\documentclass[11pt]{article}
\pdfoutput=1
\usepackage{souL}
\usepackage{jheppub}
\usepackage{amsfonts}
\usepackage{array}
\usepackage{amsmath}
\allowdisplaybreaks[4]         
\usepackage{amssymb}   
\usepackage{euscript}       
\usepackage{xcolor}          
\usepackage{tensor}     
\usepackage{caption}
\usepackage{subcaption}   
\usepackage{graphicx}
\usepackage{tikz}
\usepackage[T1]{fontenc} 
\usepackage{float}

\newcommand{\bea}{\begin{eqnarray}}
	\newcommand{\eea}{\end{eqnarray}}
\newcommand{\ba}{\begin{eqnarray}}
	\usepackage{braket}
	\newcommand{\ea}{\end{eqnarray}}

\newcommand{\beq}{\begin{equation}}
	\newcommand{\eeq}{\end{equation} }
\newcommand{\beqa}{\begin{eqnarray}}
	
	\newcommand{\eeqa}{\end{eqnarray}}
\newcommand{\beqar}{\begin{eqnarray*}}
	\newcommand{\eeqar}{\end{eqnarray*}}

\newcommand{\be}{\begin{equation}}
	\newcommand{\ee}{\end{equation}}

\newcommand{\E}{\mathcal{E}}

\renewcommand{\c}{$c$}

\allowdisplaybreaks

\title{Accelerating black holes in higher-derivative gravity}

\author[a]{Pablo A. Cano}
\author[b,c,d]{, Marina David}
\author[c]{and Simen Jacobs}

\affiliation[a]{Departamento de Física, Universidad de Murcia, Campus de Espinardo, 30100 Murcia, Spain}

\affiliation[b]{Theoretical Physics Department, CERN, 1211 Geneva 23, Switzerland}

\affiliation[c]{Instituut voor Theoretische Fysica, KU Leuven,
	Celestijnenlaan 200D, B-3001 Leuven, Belgium \vspace{0.1cm}}
\affiliation[d]{Leuven Gravity Institute, KU Leuven, Celestijnenlaan 200D, B-3001 Leuven, Belgium}

\emailAdd{pablocano@um.es}
\emailAdd{marina.claudia.david@cern.ch}
\emailAdd{simen.jacobs@kuleuven.be}

\date{\today}

\abstract{Accelerating black holes in General Relativity are described by the C-metric. We study how this geometry is modified by higher-derivative corrections, focusing on cubic curvature terms of even and odd parity. Using a suitable metric ansatz, we obtain closed-form solutions at first order in the coupling constants. The corrected geometries contain new integration constants, which correspond to physical deformations of the C-metric, and exhibit different behavior depending on the parity of the correction. In particular, parity-violating terms generate an off-diagonal metric component and induce rotation. For AdS solutions, we analyze the thermodynamics, including the Hawking temperature, entropy, and string tensions, and discuss the subtleties in defining the mass and thermodynamic string lengths.}

\begin{document} 
	\preprint{CERN-TH-2026-236}
	\maketitle
	\flushbottom
	
	\newpage
	\allowdisplaybreaks
	
\section{Introduction}

The C-metric is an exact solution of General Relativity (GR), first discovered around 1918 \cite{levi-civita:1918, Weyl:1917gp}, and named as such in the classification of \cite{Ehlers:1962zz}. The physical interpretation of this solution was not well understood until Kinnersley and Walker \cite{Kinnersley:1970zw} (see also \cite{Ashtekar:1981ar}) studied the analytically extended solution and found that it represents a pair of black holes accelerating away from each other due to the presence of strings or struts that are realized by conical singularities along the axis of acceleration. The solution can be generalized to include NUT charge, rotation and a nonzero cosmological constant \cite{Plebanski:1976gy}. Electric and magnetic charges can also be included, sourced by Maxwell theory \cite{Plebanski:1976gy} or different theories of nonlinear electrodynamics \cite{Hale:2023dpf, Hale:2025veb, EslamPanah:2026iux}. Generalizations of the C-metric beyond four dimensions are not straightforward to find. Accelerating black holes and point particles in three-dimensions were obtained and studied in \cite{Anber:2008zz, Xu:2011vp, Astorino:2011mw, Arenas-Henriquez:2022www, Cisterna:2023qhh}, while higher-dimensional C-metric-like solutions are not yet known.

The C-metric has a wealth of applications in classical and semiclassical gravity. For instance, it can be used to describe pair creation of black holes from cosmic strings \cite{Emparan:1995je, Emparan:1995eb, Hawking:1995zn,Mann:1996gj,Booth:1998gf, Dias:2003st, Dias:2004rz}. In the context of worldbrane holography, the AdS C-metric may be used to find quantum corrections to three dimensional black holes  \cite{Emparan:1999fd, Emparan:2000fn, Gregory:2008br} providing an exact solution of the semiclassical Einstein field equations known as the quantum BTZ black hole \cite{Emparan:2020znc, Climent:2024nuj,Bhattacharya:2025tdn}. 

More broadly, the C-metric allows us to learn about the impact of acceleration on black hole physics.  For example,  the thermodynamics of accelerating black holes has recently been described \cite{Astorino:2016ybm, Appels:2016uha, Appels:2017xoe,Gregory:2017ogk, Anabalon:2018qfv, Anabalon:2018ydc, Kim:2023ncn}, including its holographic interpretation, \textit{e.g.} \cite{Hubeny:2009kz, Astorino:2016xiy,Arenas-Henriquez:2023hur, Arenas-Henriquez:2025rpt}, and the case of supersymmetric generalizations of the C-metric \cite{Lu:2014sza,Ferrero:2020twa,Cassani:2021dwa, Ferrero:2021ovq}. These studies reveal that acceleration introduces several novel features that make the thermodynamic analysis considerably more subtle than for non-accelerating black holes.

The first issue that arises when defining the thermodynamics of the asymptotically flat C-metric is that the solution contains more than one horizon; aside from the black hole horizon, there is also an acceleration horizon. In these cases, the thermodynamics is not well defined as the system does not admit thermodynamic equilibrium.\footnote{However, one may consider the approach of \cite{Ball:2020vzo,Ball:2021xwt} which formulated the first law with respect to boost time. We will not consider this here but instead consider the slowly-accelerating regime ($AL<1$).} A similar issue arises for de Sitter black holes, where the second horizon is the cosmological horizon. This is an important motivation for considering the C-metric in AdS. For large acceleration, the AdS C-metric has the same horizon structure as in flat space; however, when the acceleration $A$ is smaller than the inverse of the AdS radius $L$, namely $AL<1$, the acceleration horizon is pushed beyond asymptotic infinity \cite{Podolsky:2002nk, Dias:2002mi, Krtous:2005ej}. In this regime, called the slowly-accelerating regime, there is only one horizon and equilibrium thermodynamics can be defined. 
Even in this case, there are additional subtleties related to boundary conditions. The geometry has a rather involved conformal boundary that is dependent on the angular coordinate and on the parameters of the solution and consequently, it is only locally asymptotically AdS. Therefore, it is not clear how to define an asymptotic observer and normalize the time coordinate, introducing some ambiguities in the thermodynamic description \cite{Anabalon:2018ydc,Anabalon:2018qfv,Kim:2023ncn}. 
On the other hand, the main feature of acceleration is that, due to the presence of the cosmic strings, the first law of thermodynamics must be modified to include the string tensions and their conjugate potentials, \textit{i.e.}, the thermodynamic string lengths \cite{Appels:2017xoe, Anabalon:2018qfv}.

Despite their interesting features and wide range of applications, there has been little work on the construction of accelerating black hole solutions beyond GR. For instance, generalizations of the C-metric in matter-coupled conformal gravity (without an Einstein-Hilbert term) have been studied in \cite{Meng:2016gyt,Lim:2016lxk}, while Refs.~\cite{Zhang:2019vpf,Cisterna:2021xxq} consider theories where the C-metric can be embedded --- see also \cite{Suryaatmadja:2026ais} for the case of modifications of the three-dimensional C-metric.  However, genuine higher-derivative corrections to the C-metric --- those due to corrections to the Einstein-Hilbert action containing powers of the Riemann tensor --- have not been studied yet.  The goal of this paper is to obtain accelerating black hole solutions with those corrections and to analyze some of their features. 

We focus on an effective field theory extension of GR with a nonzero cosmological constant, where the first non-trivial corrections are introduced at six derivatives. At this order, there are two cubic-curvature contributions, one of which is an even-parity term and the other a parity-violating term. Finding solutions in higher-derivative theories of gravity, whether through analytic or numerical means, is technically challenging. However, a careful choice of the metric ansatz can significantly simplify our efforts. By working with the ``polynomial'' form of the C-metric introduced in \cite{Hong:2003gx} and introducing additional polynomial form factors, we are able to obtain closed-form solutions for the first-order corrections to the C-metric. Interestingly, the even- and odd-parity corrections to the C-metric have very different forms, with the latter generating an off-diagonal component in the metric analogous to rotation. In addition, the solutions that we find are not unique: even with our polynomial ansatz the solution contains free parameters that one can relate to the choice of boundary conditions. As the C-metric has exotic asymptotic behavior, we argue that there is not a single clear way to fix these constants.

This paper is organized as follows. In Section~\ref{sec:review}, we review the C-metric in GR with a cosmological constant and its main features. In Section~\ref{sec:HD}, we consider the effective field theory extension of GR with six-derivative corrections, introduce an ansatz for the modified C-metric and obtain exact solutions of the modified Einstein field equations. We express the solutions as the sum of a particular solution plus a ``homogeneous solution'' that contains free parameters. After carefully examining the gauge redundancies in our ansatz, we conclude that some of these parameters are indeed physical deformations of the C-metric. 
In Section~\ref{sec:thermodynamics}, we obtain some of the thermodynamic properties of the corrected C-metric (temperature, entropy, string tensions) showing that the zeroth law of black hole mechanics is satisfied and the string tensions are constant. Obtaining the mass, the thermodynamic string lengths and the first law is more challenging and we discuss how this should be accomplished. 
We conclude in  Section~\ref{sec:conclusions} where we summarize our findings and provide further remarks on the subtleties of  accelerating black hole solutions beyond GR.

\section{The C-metric: a review}\label{sec:review}
Let us consider General Relativity with a cosmological constant, 
\begin{equation}
S=\frac{1}{16\pi G_{N}}\int d^{4}x\sqrt{|g|}\left(R-2\Lambda\right)\, .
\end{equation}
The C-metric is a solution of the vacuum Einstein's equations 
\begin{equation}
G_{\mu\nu}+\Lambda g_{\mu\nu}=0\, ,
\end{equation}
and it is represented by the line element
\begin{align} \label{eq:metricyx}
	ds^2 = \frac{1}{\Omega(y,x)^2}\left(-F(y) d\tau^2 + \frac{dy^2}{F(y)}+\frac{dx^2}{G(x)}+G(x) dz^2 \right)\,,
\end{align}
where
\begin{align}
	F(y) &= -\left(1-y^2\right) (1-2 A m y)-\frac{\Lambda}{3A^2}\,, \\
	G(x) &= \left(1-x^2\right) (1+2 A m x)\,,\\
\Omega(y,x) &= A (x+y)\,.
\end{align}
As we review below, this solution represents one or two (depending on the case) accelerating black holes with mass proportional to the parameter $m$ and acceleration controlled by $A$.  We will focus on the cases in which the cosmological constant is zero or negative, so from now on we set 
\begin{equation}
\Lambda=-\frac{3}{L^2}\, ,
\end{equation}
where $L$ represents the radius of AdS, and we recover the asymptotically flat case by taking $L\to \infty$. 

We have chosen to write the metric in such a way that the root structure of the functions $F(y)$ and $G(x)$ becomes explicit \cite{Hong:2003gx}. Moreover, in these coordinates $F(y)$ and $G(x)$ are related to each other via 
\begin{equation}
F(y)=-G(-y)+\frac{1}{A^2L^2}\, .
\end{equation} 
When the cosmological constant is turned off, this implies that that the metric is symmetric under the double Wick rotation given by
\begin{equation}\label{xysym}
(\tau, x, y, z)\to (i z,-y, -x, i \tau)\,.
\end{equation}
Thus, we refer to this relation when $\Lambda=0$ as the ``$xy$-symmetry'' of the C-metric.

The vanishing of the conformal factor $\Omega$ determines the location of the conformal boundary and this happens for $x+y=0$. Therefore, we constrain the coordinates $x$ and $y$ via the inequality $x+y>0$. Furthermore, to ensure the Lorentzian signature of the metric, we restrict $x$ to $ -1 \leq x \leq 1$ and assume that $2 |A| m < 1$.  This is the choice of coordinate ranges and parameters for which the C-metric can be interpreted as an accelerating black hole \cite{Podolsky:2002nk,Dias:2002mi,Griffiths:2006tk}.

\subsection{Spherical coordinates} \label{subsection:sphericalcoordinates}
The metric \eqref{eq:metricyx} takes a polynomial form in $y$ and $x$ and its structure allows us to study the solution with computational ease. However, to make the physical interpretation of the solution manifest, it is convenient to turn to another coordinate system via the transformation
\begin{align} \label{Spherictransf}
	\tau \rightarrow \frac{A t}{\alpha}\,, \quad y \rightarrow \frac{1}{A r}\,, \quad x \rightarrow \cos \theta\,, \quad z \rightarrow \frac{\phi}{K}\,,
\end{align}
where $\alpha$ and $K$ are constants. 
The line element in these Boyer-Lindquist-like coordinates then reads
\begin{align} \label{eq:merticrtheta}
	d s^2=\frac{1}{\omega(r, \theta)^2}\left[-\mathcal{F}(r) \frac{d t^2}{\alpha^2}+\frac{d r^2}{\mathcal{F}(r)}+r^2\left(\frac{d \theta^2}{\mathcal{G}(\theta)}+ \mathcal{G}(\theta) \sin ^2 \theta \frac{d \phi^2}{K^2}\right)\right]\,,
\end{align}
with
\begin{align}
	\mathcal{F}(r) & =\left(1-A^2 r^2\right)\left(1-\frac{2 m}{r}\right) +\frac{r^2}{L^2}\, ,\\
	\mathcal{G}(\theta) & =1+2 A m \cos \theta\, ,\\
	\omega(r, \theta) &=1+A r \cos \theta \,.
\end{align}
The angular coordinate $\theta$ takes values in the interval $\theta\in [0,\pi]$ while $\phi$ is $2\pi$-periodic.  The constant $K$ is related to the conical deficits at $\theta=0, \pi$, as we discuss below, while $\alpha$ fixes the normalization of the time coordinate. The choice of $\alpha$ is a subtle question that is relevant to correctly describe the thermodynamics of accelerating black holes in AdS \cite{Anabalon:2018qfv,Anabalon:2018ydc,Cassani:2021dwa,Kim:2023ncn} --- we come back to this in Section~\ref{sec:thermodynamics}.

In these coordinates, we can see that the metric smoothly reduces to the Schwarzschild-AdS black hole when we take the limit $A\to 0$ and set $K=1$. However, when $A\neq 0$ there are important features that appear. 
First of all, the asymptotic region is not located at $r\to\infty$, but at the point where the conformal factor vanishes, \textit{i.e.}, 
\begin{equation}
r=-\frac{1}{A\cos\theta}\, .
\end{equation}
On the other hand, the roots of $\mathcal{F}(r)$ represent the presence of horizons. In the case of zero cosmological constant, there are in general three horizons: $r_{+}=2m$ and $r=\pm A$. The former is a black hole horizon while the latter two represent acceleration horizons. The global interpretation of the solution is thus two black holes accelerating in the axis of symmetry, each separated by an acceleration horizon. In the AdS case, acceleration horizons still exist when the acceleration is large enough $AL > 1$.  However, in the slowly-accelerating case, $AL<1$, the acceleration horizons disappear (one can check that $\mathcal{F}(r)$ has a single real root $r_{+}$) and the spacetime contains a single black hole accelerating towards the AdS boundary. 

\begin{figure}
	\centering
	\begin{subfigure}{.33\textwidth}
		\centering
		\includegraphics[width=.7\linewidth]{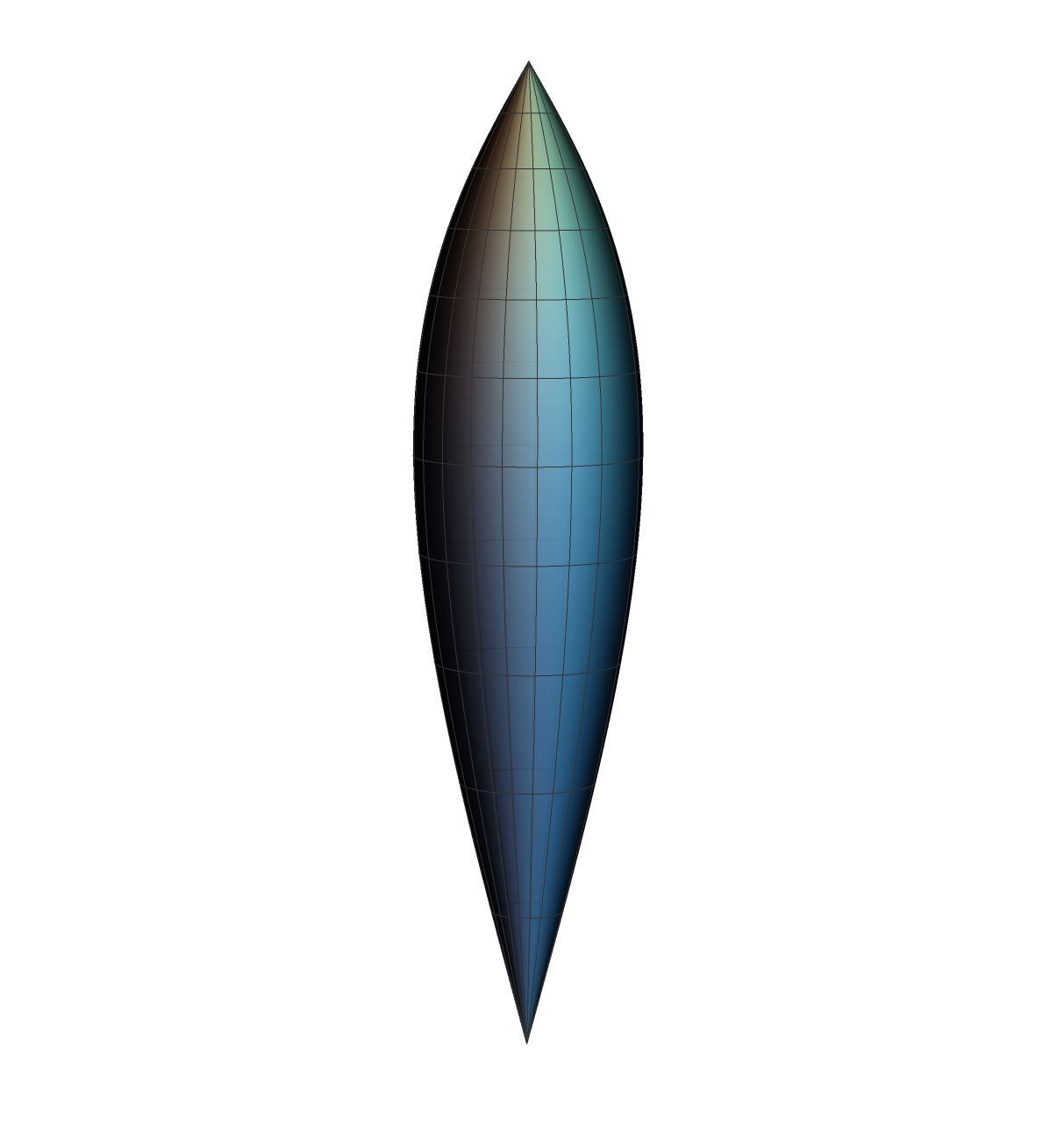}
		\caption{$K=2(1+2Am)$}
	\end{subfigure}%
	\begin{subfigure}{.33\textwidth}
		\centering
		\includegraphics[width=.7\linewidth]{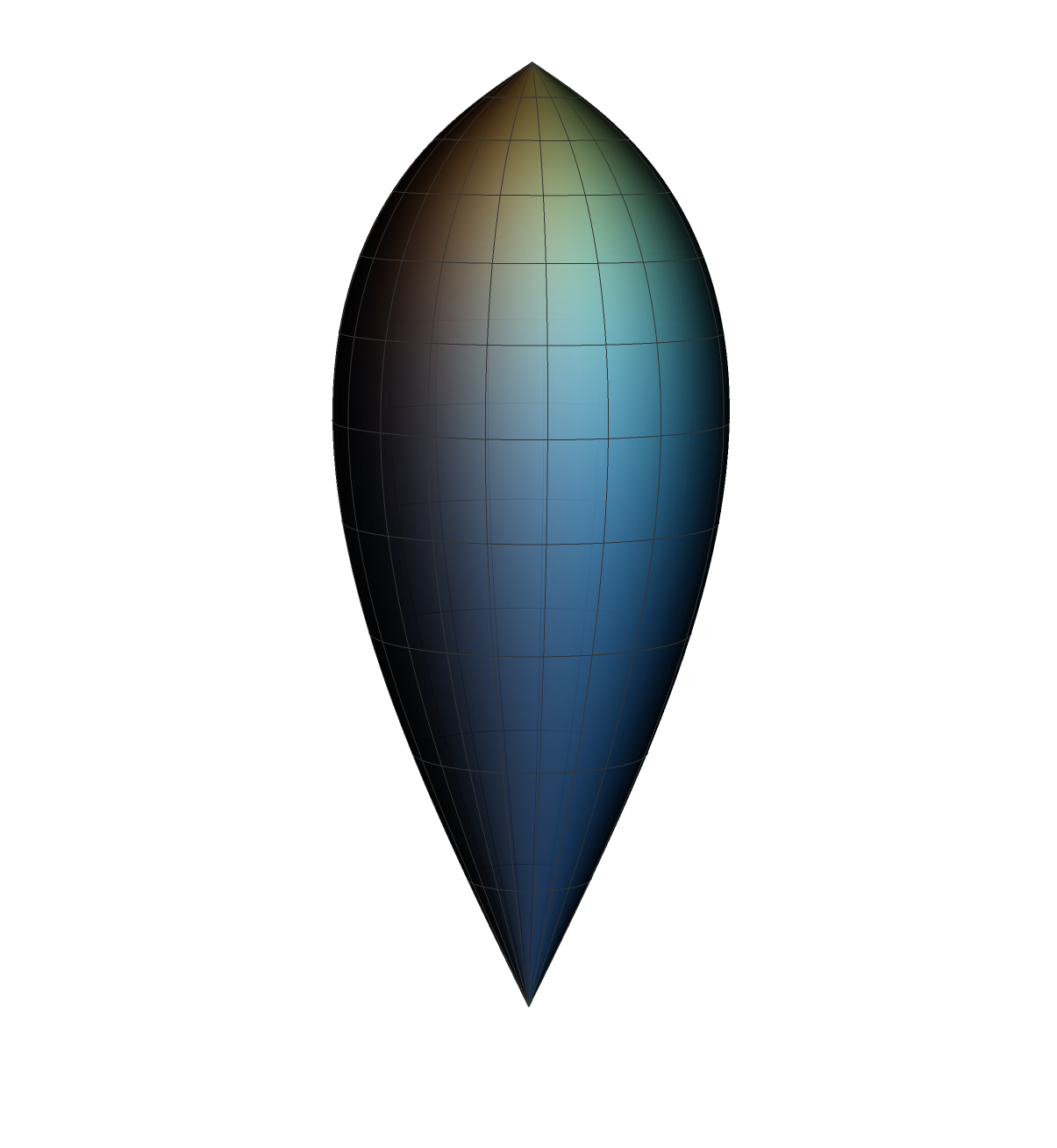}
		\caption{$K=\frac{6}{5}(1+2Am)$}
	\end{subfigure}%
	\begin{subfigure}{.33\textwidth}
		\centering
		\includegraphics[width=.7\linewidth]{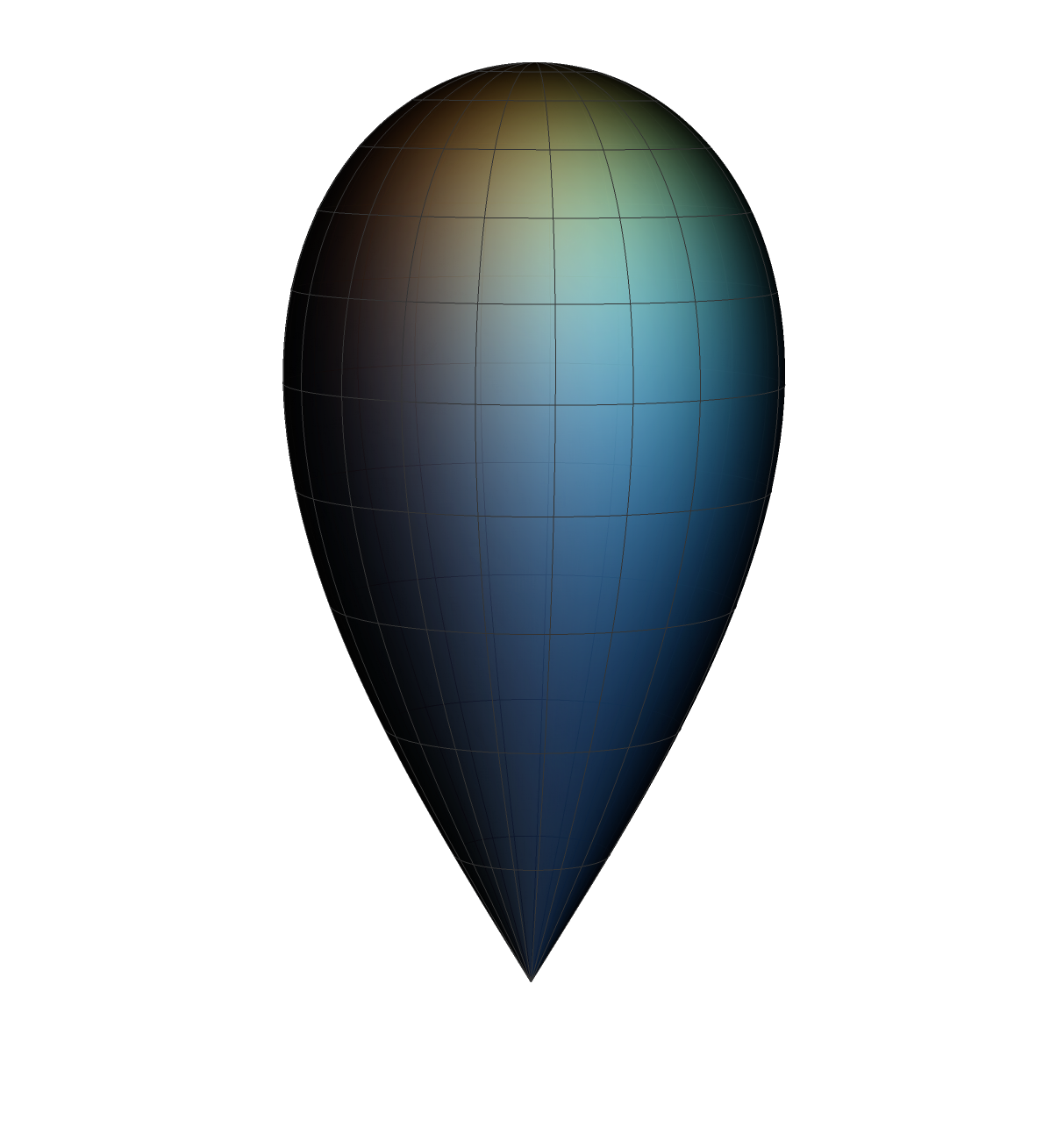}
		\caption{$K=1+2Am$}
	\end{subfigure}
	\caption{$K$ controls the conical deficits at the poles. We show several values of $K$ including the special case where the North pole does not have a deficit. As the value of $K$ increases, the horizon geometry becomes more elongated. We set the parameters to $A=0.8, m=0.2$ and $L=1$.}
	\label{fig:conicaldeficits}
\end{figure}

Another crucial aspect of the C-metric, and the counterpart of acceleration, is the existence of conical deficits on the axis of acceleration. The existence of at least one such defect extending from one of the poles of the black hole to the asymptotic boundary is unavoidable, and it is the reason why we introduced the constant $K$.  To see this more clearly, we consider a small circle at either the north ($\theta_+ = 0$) or south pole ($\theta_- = \pi$). Near the axes, the angular part of the metric reads
\begin{align}
	ds^2  \approx  \frac{r^2}{\mathcal{G}(\theta_\pm)} \left( d\theta^2 + \mathcal{G}(\theta_\pm)^2(\theta-\theta_\pm)^2  \frac{d\phi^2}{K^2} \right) + \dots \,.
\end{align}
The ratio between the circumference and the radius of the circle is not in general $2\pi$ but is given by
\begin{align}
	\frac{\text{circumference}}{\text{radius}} = \frac{2\pi\mathcal{G}(\theta_\pm)|\theta - \theta_\pm|/K}{|\theta - \theta_\pm|} = \frac{2 \pi (1 \pm 2 m A)}{K}\,.
\end{align}
The difference between this quantity and $2\pi$ is the deficit angle
\begin{align} \label{eq:GRdeficit}
	\delta_{\pm,\mathrm{GR}} =  2\pi \left(1-\frac{1\pm 2 m A}{K}\right)\,,
\end{align}
which signals the presence of a conical defect whenever $\delta_{\pm}\neq 0$. 
For $A \to 0$, one naturally chooses $K=1$ to avoid any defects in the geometry. However, for nonzero acceleration, we cannot remove both defects simultaneously. One possibility is to remove the singularity either at the North or South pole, by choosing $K = 1 + 2 m A$ or $K = 1 - 2 m A$, respectively. This can be visualized by performing an isometric embedding of the horizon into Euclidean three-dimensional space \cite{Smarr:1973zz}, as shown in Figure~\ref{fig:conicaldeficits}. If $A L > 1$, removing the deficit at the North pole leads to a metric that has the interpretation of two black holes that are connected by a strut that pushes them apart. Removing the deficit at the South pole leads to two black holes that are being pulled apart by two cosmic strings that connect each black hole to infinity. Any other value of $K$ produces both types of defects at the same time, with different tensions.
If $A L < 1$, the metric represents a single black hole, that, depending on the value of $K$, is being pulled by a cosmic string from the North pole, pushed by a cosmic string at the South pole, or both. Physically, the difference between the string tensions at each of the poles is what drives the acceleration of the black hole \cite{Podolsky:2002nk, Dias:2002mi, Krtous:2005ej}. The thermodynamics of this solution is quite rich, which we examine in Section~\ref{sec:thermodynamics}.

\section{Higher-derivative corrections to the C-metric}\label{sec:HD}

We consider a general six-derivative EFT extension of GR, which can be written as
\begin{align}\label{EFT}
	S = \frac{1}{16\pi G_{N}} \int d^{4}x\sqrt{|g|} \left(R-2\Lambda + \lambda_{\text{ev}} R_{\mu \nu}{ }^{\rho \sigma} R_{\rho \sigma}{ }^{\delta \gamma} R_{\delta \gamma}{ }^{\mu \nu}+\lambda_{\text{odd}} R_{\mu \nu}{ }^{\rho \sigma} R_{\rho \sigma}{ }^{\delta \gamma} \tilde{R}_{\delta \gamma}{ }^{\mu \nu}\right)\,,
\end{align}
where
\begin{align}\label{dualRiem}
	\tilde{R}^{\mu \nu \rho \sigma}=\frac{1}{2} \epsilon^{\mu \nu \alpha \beta} R_{\alpha \beta}{ }^{\rho \sigma}\,
\end{align}
is the dual of the Riemann tensor. In the zero-cosmological constant case, these two cubic terms represent the most general diffeomorphism-invariant modification of GR up to six derivatives, modulo redefinitions of the metric tensor. In the case of $\Lambda\neq 0$, these are again the only higher-derivative terms that cannot be removed via redefinitions of the metric and rescalings of the Newton's constant and the cosmological constant. Therefore, these cubic terms always represent the leading non-trivial correction to GR. 

The equations of motion are given by
\begin{align}\label{EFE}
G_{\mu\nu} +\Lambda g_{\mu\nu}+\E_{\mu\nu}=0\,,
\end{align}
where the effect of the corrections is expressed in the tensor $\E_{\mu\nu}$ that reads
\begin{align}
	\E_{\mu \nu}=\tensor{P}{_{(\mu}^{\rho \sigma \gamma}} R_{\nu) \rho \sigma \gamma}-\frac{1}{2} g_{\mu \nu} \mathcal{L}_{6}+2 \nabla^\sigma \nabla^\rho P_{(\mu|\sigma| \nu) \rho}\, , 
\end{align}
where 
\begin{equation}
\mathcal{L}_{6}= \lambda_{\text{ev}} R_{\mu \nu}{ }^{\rho \sigma} R_{\rho \sigma}{ }^{\delta \gamma} R_{\delta \gamma}{ }^{\mu \nu}+\lambda_{\text{odd}} R_{\mu \nu}{ }^{\rho \sigma} R_{\rho \sigma}{ }^{\delta \gamma} \tilde{R}_{\delta \gamma}{ }^{\mu \nu}
\end{equation}
is the six-derivative part of the Lagrangian and $P_{\mu\nu\alpha\beta}$ is its derivative with respect to the Riemann tensor, 
\begin{equation}\label{defofP1}
	P_{\mu\nu\alpha\beta}=\frac{\partial \mathcal{L}_{6}}{\partial R^{\mu\nu\alpha\beta}}=3 \lambda_{\mathrm{ev}} R_{\mu \nu}{}^{\alpha \beta} R_{\alpha \beta \rho \sigma}+\lambda_{\text {odd }}\left(R_{\mu \nu}{ }^{\alpha \beta} \tilde{R}_{\alpha \beta \rho \sigma}+R_{\mu \nu}{ }^{\alpha \beta} \tilde{R}_{\rho \sigma \alpha \beta}+R_{\rho\sigma}{ }^{\alpha \beta} \tilde{R}_{\mu\nu \alpha \beta}\right)\,.
\end{equation}
While the interaction proportional to $\lambda_{\text{ev}}$ preserves parity symmetry, the interaction controlled by $\lambda_{\text{odd}}$ violates parity. The C-metric is modified differently in each of these cases, and thus we discuss them separately.

\subsection*{On the structure of the corrections}
In general, we consider a perturbative expansion of the metric $g_{\mu\nu}$ of the form 
\begin{equation}
g_{\mu\nu}=g_{\mu\nu}^{(0)}+g_{\mu\nu}^{(1)}+\ldots\, ,
\end{equation}
where $g_{\mu\nu}^{(0)}$ represents a GR solution (in our case the C-metric \eqref{eq:metricyx}), and $g_{\mu\nu}^{(1)}$ a correction at first order in the coupling constants $\lambda_{\rm ev, odd}$. If we perform this expansion in the modified Einstein equations \eqref{EFE}, then we see that $g_{\mu\nu}^{(1)}$ satisfies the equation
\begin{align}\label{EFE2}
\delta G_{\mu\nu}[g^{(1)}] +\Lambda g^{(1)}_{\mu\nu}=-\E_{\mu\nu}[g^{(0)}]\,, 
\end{align}
where $\delta G_{\mu\nu}[g^{(1)}]$ is the linearized Einstein tensor evaluated on $g^{(1)}_{\mu\nu}$ and $\E_{\mu\nu}[g^{(0)}]$ is the higher-derivative tensor evaluated on the GR metric. Therefore, $g^{(1)}_{\mu\nu}$ simply satisfies the linearized Einstein equations with a source term.  Since this is a linear inhomogeneous equation, its general solution can be written as 
\begin{equation}
g^{(1)}_{\mu\nu}=g^{(1),p}_{\mu\nu}+g^{(1),h}_{\mu\nu}\, ,
\end{equation}
where $g^{(1),p}_{\mu\nu}$ is a particular solution and $g^{(1),h}_{\mu\nu}$ represents a solution of the associated homogeneous equation --- the linearized Einstein's equations in the vacuum. Thus, a possible strategy to find the solutions could be first to obtain a particular solution by following a suitable guiding principle (for instance, demanding it to be analytic or simple), and then adding a solution of the homogeneous equations in order to enforce boundary conditions. In the case of asymptotically flat or asymptotically AdS spacetimes, this often leads to a unique higher-derivative-corrected solution. In the case of the C-metric, things are more subtle, as this metric is not asymptotically flat nor asymptotically AdS (in fact, the boundary metric in the $AL<1$ case is rather involved and it depends on the parameters of the solution). This is because the C-metric is not found by imposing certain boundary conditions; instead, its defining property is that it is a Petrov type D spacetime, belonging to the family of Plebanski-Demianski spacetimes \cite{Plebanski:1976gy}. Therefore, we can think of it as a simple and analytic metric that describes accelerating black holes, but the counterpart of this is that its asymptotic structure is very involved. Conversely, there probably exists other metrics describing accelerating black holes in AdS with standard asymptotics, but these are likely not analytic and potentially only accessible numerically. 

On account of all these issues, there is no clear criterion to choose how to modify the C-metric when higher-derivative corrections are included. For instance, in the AdS case with $AL<1$, one could try to fix the boundary metric to remain the same as the one of the original C-metric. However, since the form of the boundary metric is not the guiding principle behind the construction of the C-metric, one cannot make the case very strongly for this choice. In addition, just like in GR, an analytic solution for an accelerating black hole metric with a fixed boundary geometry is unlikely to exist. On the other hand, one could try to constrain the space of solutions by demanding that the metric remains a Petrov type D spacetime, but this strategy is almost certainly hopeless in the presence of higher-derivative corrections.  

Therefore, our guiding principle will be to look for closed-form analytic solutions. As we are not fixing the form of the boundary metric, uniqueness of solutions is not granted. We could therefore have different solutions that deform the boundary metric in different ways. As we show below, this is exactly what happens. In any case, these will represent examples of accelerating black holes in higher-derivative gravity.

\subsection{Even-parity corrections}
It is more convenient to work with the $(\tau,y,x,z)$ coordinates \eqref{eq:metricyx}, in which the C-metric looks more symmetric. A general enough ansatz to include even-parity corrections is given by the following deformation of \eqref{eq:metricyx}, 
\begin{equation}\label{evenansatz}
	\begin{aligned}
		d s^2=\frac{1}{\Omega(y,x)^2} & \left\{-\left(1+\lambda_{\rm ev} f_1(y, x)\right) F(y) d \tau^2+\left(1+\lambda_{\rm ev} f_2(y, x)\right) \frac{d y^2}{F(y)}\right. \\
		& \left.+\left(1+\lambda_{\rm ev} f_3(y, x)\right) \frac{d x^2}{G(x)}+\left(1+\lambda_{\rm ev} f_4(y, x)\right) G(x) d z^2\right\}\, ,
	\end{aligned}
\end{equation}
where we include four arbitrary functions $f_{k}$, $k=1,2,3,4$, and we work perturbatively at first order in $\lambda_{\rm ev}$. We are not including any corrections in the conformal factor, which we are fixing to $\Omega(y,x)=A(x+y)$, as any modification of this function is equivalent to a redefinition of the four functions above.  On the other hand, we could also include off-diagonal terms like $d\tau dz$ and $dx dy$ in the ansatz. We are choosing not to include the former, which would represent adding rotation to the solution, and the absence of the latter is simply a gauge choice.

Our ansatz nevertheless still contains gauge freedom. To see this, we consider a perturbative coordinate transformation
\begin{align} \label{eq:coordchange}
	\tau \to (1 + \lambda_{\rm ev} c_\tau)\tau \, , \quad y \to  y + \lambda_{\rm ev}  P(y,x)\,, \quad x \to x + \lambda_{\rm ev}  Q(y,x)\,, \quad z \to (1 + \lambda_{\rm ev} c_{z})z \, ,
\end{align}
where $c_{\tau}$ and $c_{z}$ are constants and $P$ and $Q$ are functions of $y$ and $x$. 
This transformation in general introduces an off-diagonal term $dxdy$, so to remove it and to preserve the form of our ansatz we impose the constraint
\begin{align} \label{eq:crosstermeq}
	G \partial_{x}P + F \partial_{y}Q= 0\, .
\end{align}
Without loss of generality, this implies that we can write the $P$ and $Q$ functions in terms of a single potential $U(y,x)$ as
\begin{align}
	P &= F \partial_{y} U\, , \quad Q = - G \partial_{x}U\, . 
\end{align}
Then, this coordinate change \eqref{eq:coordchange} implies the following transformation rule of the functions 
\begin{subequations}\label{gaugetransf}
\begin{align}\label{ftrans}
	f_1 &\to f_1  - \frac{2(F \partial_{y} U- G \partial_{x}U)}{x+y} +F'\partial_{y} U+ 2 c_{\tau}\, ,
	\\\label{htrans}
	f_2 &\to f_2 - \frac{2(F \partial_{y} U- G \partial_{x}U)}{x+y} -F'\partial_{y} U+2\partial_{y}(F \partial_{y} U)\, ,\\\label{jtrans}
	f_3 &\to f_3 - \frac{2(F \partial_{y} U- G \partial_{x}U)}{x+y} +G'\partial_{x}U-2\partial_{x}(G \partial_{x}U)\, ,
	\\\label{ktrans}
	f_4 &\to f_4 - \frac{2(F \partial_{y} U- G \partial_{x}U)}{x+y} -G'\partial_{x}U +2 c_{z}\, ,
	\end{align}
\end{subequations}
so that, besides the freedom to shift $f_1$ and $f_4$ by a constant, we have the freedom to fix one functional constraint by choosing the function $U(y,x)$. However, we will not do this yet, as wish to keep our ansatz general. 

On top of these gauge transformations, we can also consider transformations generated by shifting the integration constants of the original solution,  
\begin{align}
	m \to m + \lambda_{\rm ev} \delta m \,, \quad A \to A + \lambda_{\rm ev} \delta A \, .
\end{align}
This moves the metric within the space of solutions and has the effect 
\begin{equation}\label{dAtransf}
\begin{aligned}
	f_1 &\to f_1+ \left(-\frac{2}{A} + \frac{\partial_{A}F}{F}
	\right)\delta A + \frac{\partial_{m}F}{F} \delta m \,,
	\\
	f_2 &\to f_2 +\left(-\frac{2}{A} - \frac{\partial_{A}F}{F}\right)\delta A - \frac{\partial_{m}F}{F} \delta m \,,
	\\
	f_3 &\to f_3  +\left(-\frac{2}{A} - \frac{\partial_{A}G}{G}\right)\delta A - \frac{\partial_{m}G}{G} \delta m \,,
	\\
	f_4 &\to f_4 + \left(-\frac{2}{A} + \frac{\partial_{A}G}{G}\right)\delta A + \frac{\partial_{m}G}{G} \delta m \,,
	\end{aligned}
\end{equation}
on the functions of the ansatz. 

We demand that these functions remain regular everywhere, which in particular implies that the horizon structure and topology of the solution are preserved. Nevertheless, following our discussion above, we do not enforce any particular conditions at the boundary $x=-y$.

\subsubsection*{Polynomial ansatz}
The functions $f_k$ satisfy a very involved system of coupled partial differential equations \eqref{EFE2}, whose resolution seems unaccessible from first principles.  

In order to look for solutions, we are inspired by the corrections to the Kerr metric \cite{Cano:2019ore}, that can be expressed as series expansion in the black hole spin where each term of the series is a polynomial in $1/r$ and $x=\cos\theta$. Here, we follow a similar strategy, although, unlike in the Kerr case, we do not perform an expansion in the acceleration $A$. Instead, we directly assume a polynomial ansatz for the functions in \eqref{evenansatz}: 
\begin{equation} \label{eq:polynomialansatz}
\begin{aligned}
f_k=\sum_{i=0}^{n}\sum_{j=0}^{n-i}\hat{f}_{k,ij}y^ix^j\, .
\end{aligned}
\end{equation}
These are polynomials of degree $n$ in $x$ and $y$. Then we insert these expressions into the equations of motion \eqref{EFE2} and we solve them order by order in $x$ and $y$, finding the values of (some) of the coefficients $\hat{f}_{k,ij}$. When $n\ge 6$, we find that this process gives an exact solution of \eqref{EFE2}. Thus, we focus on $n=6$, which is the minimal degree in order to obtain a solution. 

For $n=6$, the ansatz contains $4\times 28=112$ coefficients, and we find that the equations of motion provide $96$ constraints among them.  This means that the solution contains $16$ free parameters. Now, to further analyze the solution, it is interesting to split it as the sum of a particular solution plus a solution of the homogeneous system, which is the part containing the free constants. 
To choose a particular solution we can for instance enforce that the $m\to 0$ limit corresponds to AdS in Rindler coordinates, that the $A\to 0$ limit yields the corrected Schwarzschild-AdS black hole and we additionally impose that, in the case of $\Lambda=0$, the $xy$-symmetry \eqref{xysym} still holds. The following is a solution with those properties:
\begin{align}
f_{1}^p&=-\frac{8A^6m^2}{15}\Bigl(
35x^2-75y^2-30x^4+604x^3y-192x^2y^2-440xy^3
+75x^6+510x^5y \notag\\
&\quad
+1263x^4y^2+588x^3y^3-114x^2y^4-30xy^5
\Bigr) \notag\\
&\quad
-\frac{8A^5m}{15}\Bigl(
-75x+60y-165x^3+184x^2y+141xy^2-90y^3
+363x^3y^2+270x^2y^3
\Bigr) \notag\\
&\quad
+\frac{A^4m^2}{L^2}\Bigl(
-256x^2+192x^4+64x^3y+192x^2y^2+64xy^3
\Bigr) \notag\\
&\quad
+\frac{A^3m}{L^2}\Bigl(
-\frac{592}{3}x-32y-\frac{96}{5}x^2y+16xy^2
\Bigr)
+\frac{-4+64Amx}{L^4}\, ,\\[0.4em]
f_2^{p}&=-\frac{8A^6m^2}{15}\Bigl(
35x^2-140xy+75y^2-30x^4+232x^3y-156x^2y^2-300xy^3+90y^4
+75x^6 \notag\\
&\quad
+510x^5y+1263x^4y^2+960x^3y^3-150x^2y^4-30xy^5+30y^6
\Bigr) \notag\\
&\quad
-\frac{8A^5m}{15}\Bigl(
-5x-130y+21x^3+166x^2y+71xy^2-60y^3
+177x^3y^2+288x^2y^3
\Bigr) \notag\\
&\quad
+\frac{A^4m^2}{L^2}\Bigl(
-256x^2-128xy+192x^4+64x^3y+192x^2y^2+192xy^3
\Bigr) \notag\\
&\quad
+\frac{A^3m}{L^2}\Bigl(
-96x-32y+\frac{496}{5}x^3-\frac{144}{5}x^2y-48xy^2+16y^3
\Bigr)
-\frac{4}{L^4}\, ,\\[0.4em]
f_3^{p}&=-\frac{8A^6m^2}{15}\Bigl(
75x^2-140xy+35y^2+90x^4-300x^3y-156x^2y^2+232xy^3-30y^4
+30x^6 \notag\\
&\quad
-30x^5y-150x^4y^2+960x^3y^3+1263x^2y^4+510xy^5+75y^6
\Bigr) \notag\\
&\quad
+\frac{8A^5m}{15}\Bigl(
-130x-5y-60x^3+71x^2y+166xy^2+21y^3
+288x^3y^2+177x^2y^3
\Bigr) \notag\\
&\quad
+\frac{A^4m^2}{L^2}\Bigl(
288x^2+448xy+32y^2-480x^4-576x^3y-96x^2y^2-128xy^3
\Bigr) \notag\\
&\quad
+\frac{A^3m}{L^2}\Bigl(
\frac{296}{3}x+96y-144x^3-\frac{1176}{5}x^2y+24xy^2-8y^3
\Bigr)
+\frac{-4+64Amx}{L^4}\, ,\\[0.4em]
f_4^{p}&=\frac{8A^6m^2}{15}\Bigl(
75x^2-35y^2+440x^3y+192x^2y^2-604xy^3+30y^4
+30x^5y+114x^4y^2 \notag\\
&\quad
-588x^3y^3-1263x^2y^4-510xy^5-75y^6
\Bigr) \notag\\
&\quad
+\frac{8A^5m}{15}\Bigl(
60x-75y-90x^3+141x^2y+184xy^2-165y^3
+270x^3y^2+363x^2y^3
\Bigr) \notag\\
&\quad
+\frac{A^4m^2}{L^2}\Bigl(
-96x^2+192xy+32y^2-96x^4-320x^3y-96x^2y^2-128xy^3
\Bigr) \notag\\
&\quad
+\frac{A^3m}{L^2}\Bigl(
-\frac{568}{3}x-32y+48x^3-\frac{536}{5}x^2y+24xy^2-8y^3
\Bigr)
+\frac{-4+64Amx}{L^4}\, .
\end{align}
Let us note that the $xy$-symmetry in the $\Lambda=0$ case manifests itself through
\begin{equation}
f_1(y,x)=f_4(-x,-y)\, ,\quad f_2(y,x)=f_3(-x,-y)\, .
\end{equation}

Next we analyze the ``homogeneous'' part of the solution, which depends on 16 free parameters. Naturally, many of them will be gauge parameters on account of the gauge freedom represented by \eqref{gaugetransf}. In order to determine if all of them are gauge, we have to check if we can absorb them in the function $U(y,x)$. To this end, let us take the homogeneous part of the solution $f_{k}^h$ and assume there is a transformation $U(y,x)$ that sets all these functions to zero via \eqref{gaugetransf}. As this is a system of four equations for one function $U$, a solution only exists if several integrability conditions are satisfied. From the equations \eqref{ftrans} and \eqref{ktrans}, setting the transformed $f$ and $k$ to zero,  we can obtain explicitly $\partial_{x}U$ and $\partial_{y}U$:

\begin{align}
\partial_{y}U&=-\frac{2 G f_4^h+f_1^h \left((x+y) G'-2 G\right)}{G' \left(2 F-(x+y) F'\right)+2 G F'}\, ,\\
\partial_{x}U&=\frac{2 F f_1^h+f_4^h \left((x+y) F'-2 F\right)}{G' \left(2 F-(x+y) F'\right)+2 G F'}\, .
\end{align}
A first integrability condition is therefore $\partial_{y}(\partial_{x}U)= \partial_{x}(\partial_{y}U)$. On the other hand, once we have $\partial_{x}U$ and $\partial_{y}U$, we can substitute these expressions in the equations \eqref{htrans} and \eqref{jtrans} demanding the result to be zero in order to obtain two additional constraints that must be satisfied if the solution is pure gauge. When we evaluate explicitly these integrability conditions, we see that in general they are not satisfied, and they imply four additional constraints on the coefficients of the solution. Two of these constraints are equivalent to choosing the constants $c_{\tau}$ and $c_{z}$, which are also a gauge choice. The other two constraints are physical and indicate a non-trivial deformation of the C-metric. One could think that this deformation corresponds to a shift in the acceleration $A$ and mass parameter $m$. However, as implied by \eqref{dAtransf}, this transformation produces a non-polynomial form of the $f_{k}$ functions, and we have not found a transformation that can put this in a polynomial form. Therefore, the conclusion is that the solution that we found contains two new genuine integration constants that represent physical deformations of the C-metric. 
There are of course infinitely many ways of expressing these two linearly independent solutions because of gauge redundancies. In order to express them in a simple form, we try to remove terms with higher powers of $x$ and $y$. We find there is a single solution that contains at most quadratic powers of $x$ and quadratic powers of $y$. We call this the $a$-type deformation and it is explicitly given by
\begin{align}
f_1^{a}(y,x)=&A^4\left[x-y+3A m \left(2 x^2-2 x y- y^2\right)-6 A^2 m^2 \left(x+2 y+6 x^2 y+3 x y^2\right)\right]\, ,\\
f_2^{a}(y,x)=&A^4\left[x+y-3A m \left(3-2 x^2+2 x y+y^2\right)-6 A^2 m^2 \left(x-2 y+6 x^2 y+3 x y^2\right)\right]\, ,\\
f_3^{a}(y,x)=&f_2^a(-x,-y)+\frac{9A^3m}{L^2}\,,\\
f_4^{a}(y,x)=&f_1^a(-x,-y)\, ,
\end{align}
up to a free normalization constant. We observe that this one also obeys the $xy$-symmetry in the $\Lambda=0$ case. 

The other independent solution requires to have at least cubic powers of $x$ and $y$ and a possible way to express it is 
\begin{align}\notag
f_1^{b}(y,x)=&A^4\bigg[-x+y+A m \left(-6 x^2+6 x y+3 y^2\right)+A^2 m^2 \left(-57 x-60 y-84 x^2 y-27 x y^2\right)\\\notag
&+A^3 m^3 \left(-78 x^2+108 x y-120 x^3 y+54
   y^2+90 x^2 y^2+60 x y^3\right)\\
   &+A^4 m^4 \left(-12 x-24 y-72 x^2 y-36 x
   y^2\right)+\frac{60 m^2 x}{L^2}\bigg]\, ,\\\notag
f_2^{b}(y,x)=&A^4\big[-x-y+A m \left(9-6 x^2+6 x y+3 y^2\right)+A^2 m^2 \left(3 x-84 x^2 y-87 x y^2\right)\\\notag
&+A^3 m^3 \left(-18-78 x^2-12 x y-120 x^3 y+54 y^2+90 x^2 y^2+180 x
   y^3\right)\\
   &+A^4 m^4 \left(-12
   x+24 y-72 x^2 y-36 x y^2\right)\big]\, ,\\\notag
f_3^{b}(y,x)=&A^4\bigg[x+y+A m \left(9+3 x^2+6 x y-6 y^2\right)+A^2 m^2 \left(-3 y+87 x^2 y+84 x y^2\right)\\\notag
&+A^3 m^3 \left(-18+54 x^2-12
   x y+180 x^3 y-78 y^2+90 x^2 y^2-120 x y^3\right)\\
   &+A^4 m^4 \left(-24
   x+12 y+36 x^2 y+72 x y^2\right)+\frac{-\frac{9 m}{A}+18 A m^3+15 m^2 x}{L^2}\bigg]\,,\\\notag
f_4^{b}(y,x)=&A^4\bigg[-x+y+A m \left(3 x^2+6 x y-6 y^2\right)+A^2 m^2 \left(60 x+57 y+27 x^2 y+84 x y^2\right)\\\notag
&+A^3 m^3 \left(54 x^2+108 x y+60 x^3 y-78
   y^2+90 x^2 y^2-120 x y^3\right)\\
   &+A^4 m^4 \left(24 x+12 y+36 x^2 y+72 x
   y^2\right)-\frac{75 m^2 x}{L^2}\bigg]\, ,
\end{align}
again up to a free normalization constant. We call this the $b$-type deformation. 

To sum up, we have found that, within a sixth-degree polynomial ansatz \eqref{eq:polynomialansatz} one can find closed form solutions for the corrections to the C-metric. These are not unique and are in general given by
\begin{equation} \label{eq:totalevensolution}
f_k=f_k^{p}+a f_k^{a}+b f_k^{b}\, ,
\end{equation}
where $a$ and $b$ are free dimensionless parameters.

\subsubsection*{Boundary metric}
As in the usual C-metric, the asymptotic region of the metric \eqref{evenansatz} is located at $x+y\to 0$. 
In the AdS case, we can define a conformal boundary metric, which we derive by introducing the variable $R$ given by
\begin{equation}
y=-x+\frac{1}{AR}\, ,
\end{equation}
and expanding the metric for $R\to\infty$. At leading order, we get 
\begin{equation}\label{boundarymetric1}
	\begin{aligned}
		d s^2_{\infty}&=R^2 \Bigg\{-\left(1+\lambda_{\rm ev} f_1(-x, x)\right) F(-x) d \tau^2+\left(1+\lambda_{\rm ev} f_4(-x, x)\right) G(x) d z^2\\
		&+\frac{d x^2}{(AL)^2G(x)F(-x)}\left[1+(AL)^2\lambda_{\rm ev}\left(G(x) f_2(-x, x)+F(-x) f_3(-x, x) \right)\right] \Bigg\}+\mathcal{O}(R)\, .
	\end{aligned}
\end{equation}
As the functions $f_{k}(-x,x)$ are nonzero and have a nontrivial $x$-dependence, the boundary metric is modified. However, we still have to account for gauge freedom in the choice of coordinates. For instance, we can redefine the coordinate $x\to x+\lambda_{\rm ev} \Delta(x)$ such that  $\left(1+\lambda_{\rm ev} f_1(-x, x)\right) F(-x)\to C^2 F(-x)$ for an arbitrary constant $C$, and then rescale $\tau\to C^{-1}\tau$ so that the component $g_{\tau\tau}^{\infty}$ takes the same form as in the C-metric. This fully fixes the gauge freedom in the coordinate $x$. Using this new coordinate we can then evaluate $g^{\infty}_{xx}$ and $g^{\infty}_{zz}$.  We observe that in general they receive corrections, and within the family of  solutions \eqref{eq:totalevensolution}, there is no choice of integration constants that make these corrections vanish. Thus, the solutions that we find deform the boundary metric.

\subsubsection*{Spherical coordinates}
The solutions can also be expressed in the spherical coordinates given in \eqref{Spherictransf}, yielding
\begin{equation}
\begin{aligned} \label{eq:metricrtheta2}
	d s^2=\frac{1}{\omega(r, \theta)^2}\Bigg[&-\left(1+\lambda_{\rm ev} f_1\right) \mathcal{F}(r) \frac{dt^2}{\alpha^2}+\left(1+\lambda_{\rm ev} f_2\right) \frac{d r^2}{\mathcal{F}(r)}\\
	&+r^2\left(1+\lambda_{\rm ev} f_3\right) \frac{d \theta^2}{\mathcal{G}(\theta)}+r^2 \left(1+\lambda_{\rm ev} f_4\right) \mathcal{G}(\theta) \sin ^2 \theta \frac{d \phi^2}{K^2}\Bigg]\,. 
\end{aligned}
\end{equation}
The functions $f_k$ are the same as in \eqref{evenansatz} but are expressed in terms of $x=\cos\theta$ and $y=1/(A r)$. In order to avoid introducing more notation, in Section \ref{sec:thermodynamics} below we use the abuse of notation $f_{k}(r,\theta)$ to refer to $f_{k}\left(y=(Ar)^{-1}, x=\cos\theta\right)$.

\subsubsection*{Horizon geometry}

\begin{figure}[t]
	\centering
	\begin{subfigure}[c]{0.35\textwidth}
		\centering
		\includegraphics[width=0.7\textwidth]{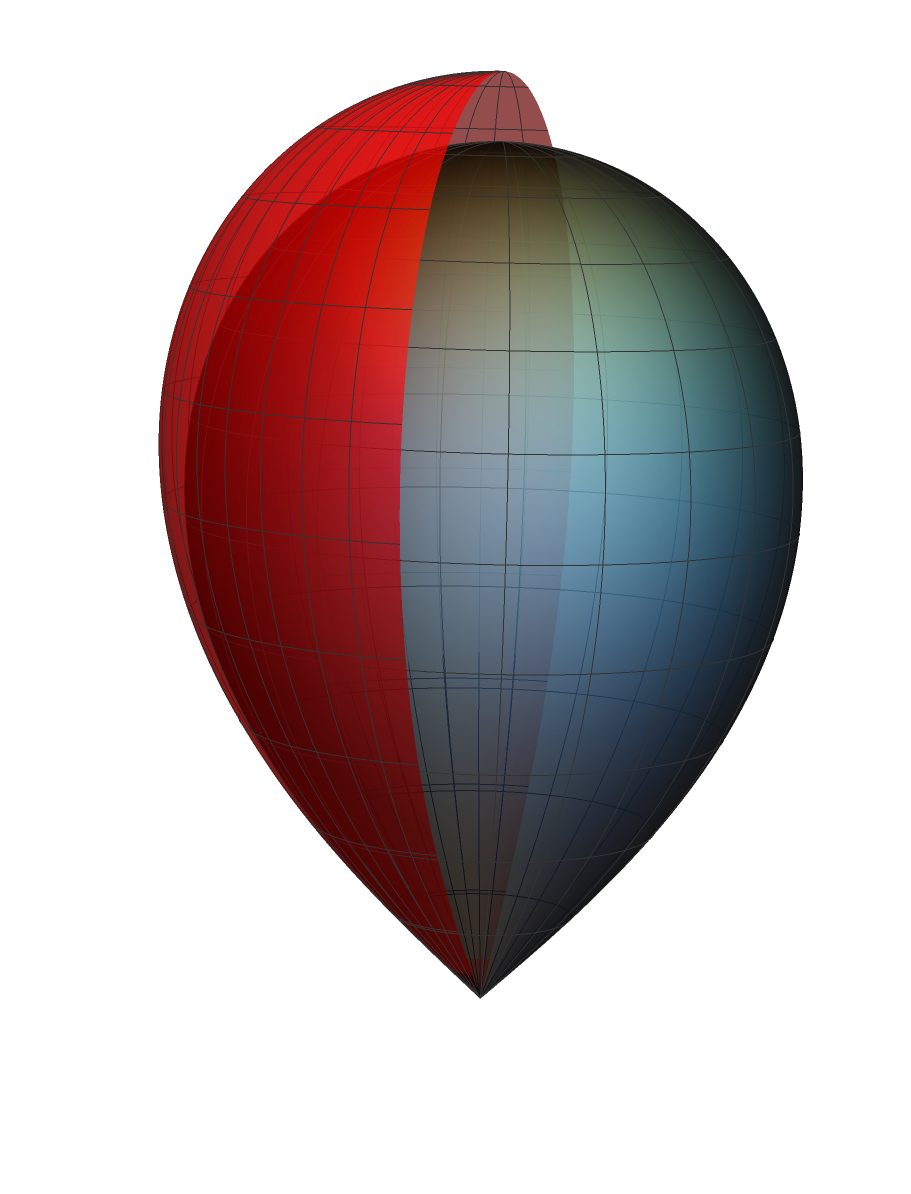}
		\caption{$A=0.2$}
	\end{subfigure}
	\hfill
	\begin{subfigure}[c]{0.35\textwidth}
		\centering
		\includegraphics[width=0.7\textwidth]{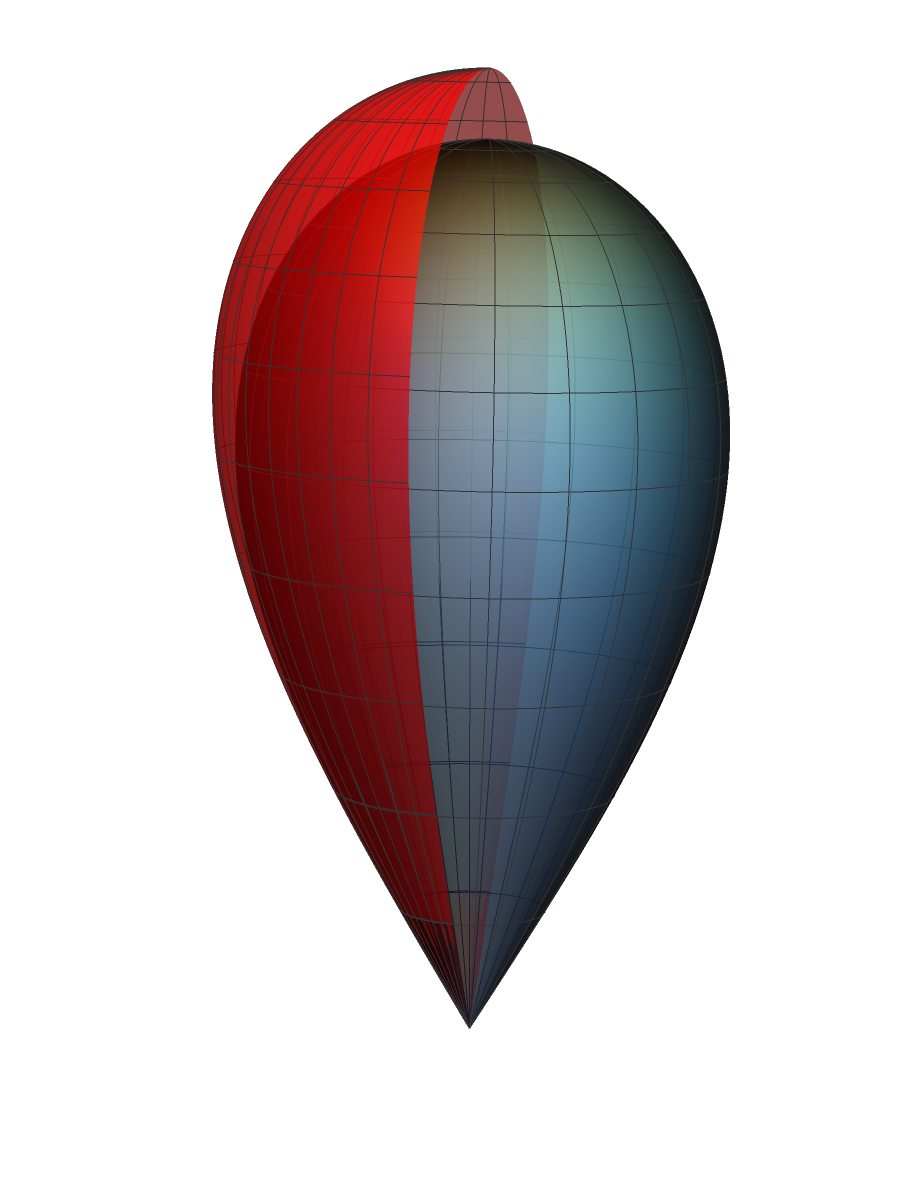}
		\caption{$A=0.4$}
	\end{subfigure}
	\hfill
	\begin{subfigure}[c]{0.2\textwidth}
		\centering
		\includegraphics[width=0.6\textwidth]{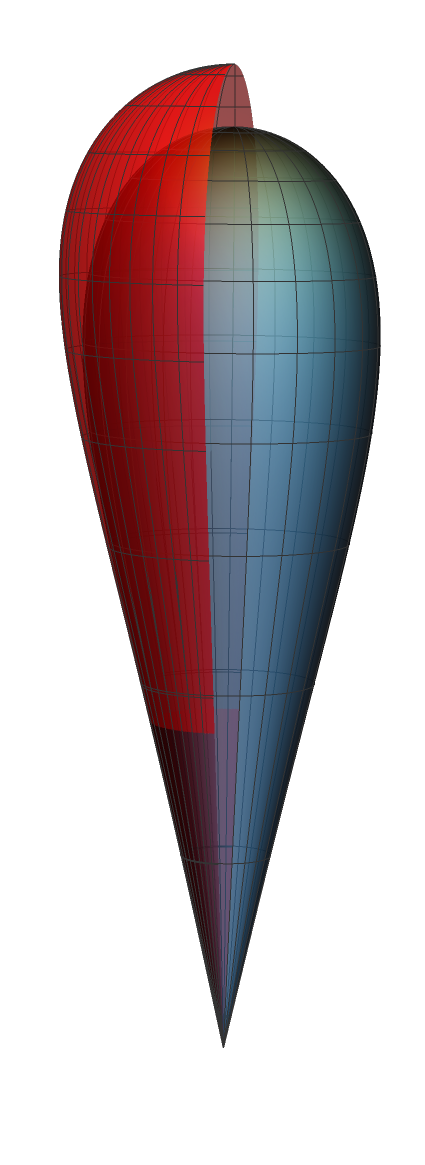}
		\caption{$A=0.8$}
	\end{subfigure}
	\hfill
	
	\caption{The horizon geometry. We set $L=1$ and $m=0.4$ and vary the acceleration. We choose $K$ so that there is no conical deficit at the North pole, even when the corrections are turned on. The  darker surface is the horizon of the GR solution while the red surface is the corrected horizon with $a=b=0$ and with a negative coupling constant $\lambda_{\mathrm{ev}}=-10^{-3}$. For (a) and (b), the entire horizon becomes larger with the largest difference being closer to the North pole. (c) For larger values of the acceleration, the higher-curvature corrections enlarge the upper hemisphere while the lower hemisphere becomes smaller.}
	\label{fig:conicaldeficitswithcorrections}
\end{figure}

The higher-derivative corrections deform the horizon geometry and this deformation relies on the values of the parameters of the solution and on the sign of the coupling constant. To explicitly show this, we plot the isometric embedding of the horizon in Euclidean space in Figure~\ref{fig:conicaldeficitswithcorrections} for both the uncorrected and corrected solutions (considering $a=b=0$ for simplicity).  
The horizon metric in $(\theta,\phi)$ coordinates can be mapped to a surface in three-dimensional Euclidean space given parametrically by \cite{Smarr:1973zz}
\begin{align}
	X = \rho(\theta)\cos \phi\,, \quad Y=\rho(\theta)\sin \phi\,, \quad Z=\int d\theta \sin\theta\sqrt{g_{\theta \theta}-(\rho'(\theta))^2}\,,
\end{align}
where
\begin{align}
	\rho(\theta) = \sqrt{g_{\phi\phi}} = \frac{r_+ \sin\theta \sqrt{\mathcal{G}(\theta)} }{K\omega(r_+,\theta)}\sqrt{1+\lambda_{\mathrm{ev}}f_4(r_{+},\theta)}\,,
\end{align}
and $(X,Y, Z)$ are Cartesian coordinates.  
We find that for a positive value of the coupling constant, the horizon geometry shrinks, with the largest difference being near the North pole. On the other hand, for a negative value of the coupling constant, the horizon bulges outward. We have chosen a value of $K$ such that North pole is smooth and only the South pole features a conical defect. For GR, this is achieved for $K=1+2Am$, but a different value of $K$ is required when we introduce the higher-derivative corrections --- see Section~\ref{sec:thermodynamics} and equation~\eqref{Ksmooth}.

\subsubsection*{Limit $m\to 0$ and map to global AdS}
It is known that the $m\to 0$ limit of the C-metric corresponds to AdS/flat spacetime in the Rindler patch \cite{Appels:2017xoe,Gregory:2017ogk}. In the case of the higher-derivative corrections, the limit $m \to 0$ of the solution \eqref{eq:totalevensolution} yields
 \begin{subequations}
\begin{align}
	f_1&=-\frac{4}{L^4}+A^4(a-b)(x-y)\, ,\\
	f_2&=-\frac{4}{L^4}+A^4(a-b)(x+y)\, ,\\
	f_3&=-\frac{4}{L^4}-A^4(a-b)(x+y)\, ,\\
	f_4&=-\frac{4}{L^4}+A^4(a-b)(x-y)\, .
	\end{align}
\end{subequations}
In order to ensure that the corresponding metric is pure AdS, we have to impose $a=b$. We observe this is a very mild condition, as one may assume that $a$ and $b$ depend on $A$ and $m$, and one would only need to impose that $a\to b$ when $m\to 0$. 
Assuming this, the limit $m\to 0$ of \eqref{eq:metricrtheta2} is found to be
\begin{align}
	d s^2=\frac{1}{\omega^2}\frac{L_{\rm{eff}}^2}{L^2}\left[-\left(1+\frac{r^2}{L^2}\left(1-A^2 L^2\right)\right) \frac{d t^2}{\alpha^2}+\frac{d r^2}{1+\frac{r^2}{L^2}\left(1-A^2 L^2\right)}+r^2\left(d \theta^2+\sin ^2 \theta \frac{d \phi^2}{K^2}\right)\right]\,,
\end{align}
where 
\begin{equation}
L_{\rm{eff}}=L - \frac{2 \lambda_{\mathrm{ev}}}{L^3}\,,
\end{equation}
is the radius of AdS, which receives corrections. In order to show that this metric is vacuum AdS, we perform the coordinate transformation \cite{Podolsky:2002nk}
\begin{align}
	1+\frac{R^2}{L^2}=\frac{1+\frac{r^2}{L^2}\left(1-A^2 L^2\right)}{\left(1-A^2 L^2\right) \omega^2}
	\,, \quad R \sin \Theta=\frac{r \sin \theta}{\omega}\,.
\end{align}
This results in the following metric
\begin{align}
	d s_{\mathrm{AdS}}^2=\frac{L_{\rm{eff}}^2}{L^2} \left\{-\left(1-A^2 L^2\right)\left(1+\frac{R^2}{L^2}\right) \frac{d t^2}{\alpha^2}+\frac{d R^2}{1+\frac{R^2}{L^2}}+R^2\left(d \Theta^2+\sin ^2 \Theta \frac{d \phi^2}{K^2}\right)\right\}\,.
\end{align}
Finally, if we let $R=\rho L/L_{\rm eff}$, we find the metric can be written in global AdS coordinates
\begin{align}
	d s_{\mathrm{AdS}}^2=-\frac{L_{\rm{eff}}^2}{L^2}\left(1-A^2 L^2\right)\left(1+\frac{\rho^2}{L_{\rm eff}^2}\right) \frac{d t^2}{\alpha^2}+\frac{d \rho^2}{1+\frac{\rho^2}{L_{\rm eff}^2}}+\rho^2\left(d \Theta^2+\sin ^2 \Theta \frac{d \phi^2}{K^2}\right)\,. 
\end{align}
From this expression, we find that the standard normalization for the time coordinate $t$ corresponds to
\begin{align}\label{alphavalue}
	 \alpha = \frac{L_{\rm{eff}}}{L}\sqrt{1-A^2 L^2}\,.
\end{align}
However, this identification of $\alpha$ is only rigorous for $m=0$.

\subsection{Odd-parity corrections}
The parity-violating cubic correction generates a right-hand-side of the equation \eqref{EFE2} that contains only one component which is off-diagonal, $\E_{\tau z}[g^{(0)}]$.
Therefore, the ansatz of the form \eqref{evenansatz} does not solve the equations of motion for the parity-violating correction, as this ansatz gives a vanishing $tz$ component in the left-hand-side of \eqref{EFE2}.  Therefore, a sufficient ansatz must include an off-diagonal  $d\tau dz$ component of the metric, while the remaining components may remain unchanged:
\begin{equation}\label{eq:oddansatz}
	\begin{aligned}
		d s^2=\frac{1}{\Omega(y,x)^2} & \left\{-F(y) d \tau^2+ \frac{d y^2}{F(y)}+\frac{d x^2}{G(x)}+ G(x) d z^2 + \lambda_{\mathrm{odd}}q(y,x)d\tau dz\right\}\, .
	\end{aligned}
\end{equation}
Physically, the off-diagonal piece plays the role of rotation, so this means that odd-parity corrections necessarily induce some kind of rotation along the axis of acceleration

The only nontrivial component of the modified Einstein equations is the $\tau z$ one, which gives us the differential equation
\begin{align} \label{eq:oddparityEoM}
	F \partial_y^2\left(q\Omega^{-1}\right)+G \partial_x^2\left(q\Omega^{-1}\right)+\left(4 A m (x+y) - \frac{2}{A^2L^2(x+y)^2}\right)\Omega^{-1}q=864 A^5 m^2 F G (x+y)^3\,.
\end{align}
In the asymptotically flat case, the homogeneous part of this equation is separable, but the full inhomogeneous equation is not. This is due to the appearance of the conformal factor to the third power. When we introduce the cosmological constant, separability is lost even at the level of the homogeneous part of the equation, as there is a $L$-dependent term that couples $x$ and $y$.

As in the even-parity case, we can try to solve the equation using a polynomial ansatz, so we consider
\begin{equation} \label{eq:qpolynomialansatz}
\begin{aligned}
q=\sum_{i=0}^{n}\sum_{j=0}^{n-i}\hat{q}_{ij}y^ix^j \, .
\end{aligned}
\end{equation}
In the asymptotically flat case, we find that the equation \eqref{eq:oddparityEoM} admits exact solutions if we set at least $n=6$, and in that case the general solution reads 
\begin{align}
	q(y,x) &= -\frac{6 A^3}{m} F(y) G(x) (x+y) \left(1-2 A m (x-y) + 4 A^2 m^2 \left(x^2+5 x y+y^2+4\right)\right)+q_{h}\, ,
\end{align}
where $q_{h}$ is a solution of  the associated homogeneous equation containing four integration constants, 
\begin{align}
	q_h &= \alpha_1 G(x) + \alpha_2 F(y) + \alpha_3 (x+y)[1-x y(1+2 A m (x-y))] \nonumber
	\\& + \alpha_4 (x+y) \left(-y^2 \left(1-x^2\right)-x (x-y) (1-2 A m y)\right)\, .
\end{align}
The constants $\alpha_1$ and $\alpha_2$ correspond to local gauge transformations $z\to z+\tfrac{1}{2}\alpha_1\lambda_{\rm odd} \tau$ and $\tau\to \tau-\tfrac{1}{2}\alpha_2\lambda_{\rm odd} z$, respectively, while $\alpha_3$ and $\alpha_4$ represent non-trivial deformations of the C-metric. 

In terms of the $t$, $\phi$ coordinates, this implies that the horizon has an angular velocity,  
\begin{equation}
\Omega_{H}=-\frac{g_{t\phi}}{g_{\phi\phi}}\bigg|_{y=y_{+}}=-\lambda_{\rm odd}\frac{q(y_{+},x) K}{2\alpha G(x)}=\frac{\lambda_{\rm odd} K}{16 \alpha  A^3 m^3} \left[\alpha_4-4 A^2 m^2 (2 A m\alpha_1+\alpha_3)\right]\, ,
\end{equation}
while the frame at infinity is also rotating, with an angular velocity 
\begin{equation}
\Omega_{\infty}=-\frac{g_{t\phi}}{g_{\phi\phi}}\bigg|_{y=-x}=-\lambda_{\rm odd}\frac{q(-x,x) K}{2\alpha G(x)}=-\lambda_{\rm odd}(\alpha_1-\alpha_2)\frac{K}{2\alpha}\, .
\end{equation}
We could for instance choose $\alpha_2=\alpha_1=-(2Am)^{-1}\alpha_3+(2Am)^{-3}\alpha_4$, so that $\Omega_{H}=\Omega_{\infty}=0$.  Even with this choice, $g_{t\phi}$ is nonvanishing at intermediate locations, implying that local observers experience frame-dragging. We also observe that even imposing $\Omega_{H}=\Omega_{\infty}=0$, the solution still contains the free parameters $\alpha_3$ and $\alpha_4$.

In the case of a non-zero cosmological constant, we did not find any polynomial ansatz that solves \eqref{eq:oddparityEoM}  exactly, even if we increase the degree of the polynomial, and we have not been able to find a closed-form solution. We leave this for future work.

\section{Thermodynamics} \label{sec:thermodynamics}

In this section, we carefully examine the thermodynamics of the accelerating black hole solutions. We work in the slowly-accelerating regime $AL<1$ where the solutions contain a single black hole in AdS and equilibrium thermodynamic quantities can be defined. 
We note that odd-parity corrections do not modify thermodynamic quantities at first order in the coupling constant $\lambda_{\rm odd}$ \cite{Cano:2019ore} and thus we focus entirely on the solution with even-parity corrections \eqref{eq:metricrtheta2}. For simplicity, we only show explicit expressions for the particular solution in \eqref{eq:totalevensolution} and neglect the a-type and b-type deformations.

In the slowly-accelerating regime $AL<1$ there is no acceleration horizon and the black hole horizon is the only real root of $\mathcal{F}(r)$.  The explicit expression of $r_{+}$ from $\mathcal{F}(r_+)=0$ is quite involved, but one can instead express $m$ as a function of $r_{+}$ as 
\begin{equation}
	m=\frac{r_{+}}{2}\left(1+\frac{r_{+}^2}{L^2(1-A^2r_{+}^2)}\right)\, .
\end{equation} 
Note that due to the ansatz of the corrected metric, the location of the horizon or equivalently, the value of $m$ in terms of $r_+$ remains the same as in GR. 
For a fixed cosmological constant, the solution is thus determined by three free parameters $\{m, A, K\}$ or equivalently $\{r_{+}, A, K\}$. A thermodynamic description of this accelerating black hole would consequently require three independent thermodynamic variables.

\subsubsection*{Temperature}
The Hawking temperature can be derived, as usual, by a Wick-rotation of the time coordinate $t_{E}=i t$ and demanding that $r=r_{+}$ is a smooth cap of the resulting Euclidean geometry. Requiring that no conical defects appear at $r=r_{+}$ fixes the periodicity $\beta$ of the Euclidean time, and the Hawking temperature is its inverse, $T=\beta^{-1}$. 
The computation is standard from the diagonal metric \eqref{eq:metricrtheta2}, and expanding to linear order in $\lambda_{\rm ev}$, we have
\begin{align}
	T =  \frac{\mathcal{F}'(r_+) \left( 1 + \frac{1}{2}\lambda_{\mathrm{ev}} (f_1(r_+,\theta)-f_2(r_+,\theta))\right)}{4\pi \alpha}\,.
\end{align}
Note that this expression could a priori depend on $\theta$. However, the $\theta$-dependence cancels exactly such that $T$ is constant over the horizon. As such, this class of solutions obeys the zeroth law of black hole mechanics. Explicitly, for the solution with $a=b=0$, we find
\begin{align}
	\frac{T}{T_{\mathrm{GR}}} = 1+\lambda_{\mathrm{ev}}A^4\left(-\frac{46}{3}+\frac{2+10 A^2 r_+^2}{A^4 r_+^4}+\frac{60-16 A^2 r_+^2}{3A^2L^2 \left(1-A^2 r_+^2\right)}+\frac{2 \left(-1+5 A^2 r_+^2+5 A^4 r_+^4\right)}{A^4L^4 \left(1-A^2 r_+^2\right){}^2}\right)\,,
\end{align}
where $T_{\mathrm{GR}}$ is the Hawking temperature of the C-metric in GR given by
\begin{equation}\label{TGR}
	T_{\mathrm{GR}} = \frac{\mathcal{F}'(r_+)}{4\pi \alpha}\,.
\end{equation}
Naturally, the temperature is inversely proportional to the parameter $\alpha$ that determines the normalization of the time coordinate at the boundary. However, how to choose this parameter is a complex problem, and we comment on it below. 

\subsection*{String Tension}
The tensions of the cosmic strings attached to the black hole are directly proportional to the conical deficits $\delta_{\pm}$ at the North and South poles (respectively $\theta_{+}=0$, $\theta_{-}=\pi$) and are given by 
\begin{align}
	\mu_\pm = \frac{\delta_\pm}{8\pi G_{N}}\,.
\end{align}
To find their values, we consider a small circle around each of the poles, with $\theta=\theta_{\pm}+\Delta\theta$ and $\phi$ performing a full $2\pi$ turn.  Using the metric \eqref{eq:metricrtheta2}, we obtain the following ratio between the radius and circumference of this circle in the limit $\Delta\theta\to 0$: 
\begin{align}
	\frac{\text{circumference}}{\text{radius}} = \frac{2\pi\mathcal{G}(\theta_\pm) \sqrt{1 + \lambda_{\mathrm{ev}} f_4(r,\theta_\pm)}}{K \sqrt{1 + \lambda_{\mathrm{ev}} f_3(r,\theta_\pm)}} \,.
\end{align}
Since we are working at linear order in the coupling constant this can be rewritten as
\begin{align}
	\delta_\pm = 2\pi \left[1- \left (\frac{1\pm 2 m A}{K} \right) \left(1+\frac{\lambda_{\mathrm{ev}}}{2}\left(f_4\left(r,\theta_\pm\right)-f_3\left(r,\theta_\pm\right) \right ) \right ) \right ]\,.
\end{align}
In contrast to the solution without higher-derivative corrections, the tensions in the expression above could a priori be $r$-dependent, meaning that the tension of the strings would vary depending on the position along the axis of acceleration. However, when we fill in our solution, the $r$-dependence drops out. This is again a non-trivial check of the consistency of the solution. Once the dust settles, the deficits are found to be
\begin{align}
	\delta_\pm = 2\pi \left[1- \left (\frac{1\pm 2 m A}{K} \right) \left(1\pm\frac{8\lambda_{\mathrm{ev}}}{3}A^5 m \left(16\pm 27 A m-\frac{18}{A^2L^2}\right) \right ) \right ]\, ,
\end{align}
for the solution with $a=b=0$. 
As in the case of GR, one can set a particular value of $K$ such that one of the defects vanishes, but it is never possible to get rid of both of them. For instance, we can get rid of the string at the North pole by setting
\begin{equation}\label{Ksmooth}
K= \left (1+ 2 m A\right) \left[1+\frac{8\lambda_{\mathrm{ev}}}{3}A^5 m \left(16+27 A m-\frac{18}{A^2L^2}\right) \right]\, ,
\end{equation}
and then the string at the South pole has a tension
\begin{equation}
\mu_{-}=\frac{A m}{G_{N}(1+2 A m)}\left[1-\frac{8 \lambda_{\rm ev}A^2}{3L^2}\left(9-8 A^2 L^2\right) (1-2 A m)\right]\, .
\end{equation}

\subsection*{Entropy} 
One can straightforwardly obtain the black hole entropy of the uncorrected C-metric from the Bekenstein-Hawking area formula, which yields
\begin{align}\label{SGR}
	S_{\mathrm{GR}} = \frac{\pi r_+^2}{G_{N}K(1-A^2 r_+^2)}\,.
\end{align}
In the presence of higher-derivatives in the action, the entropy is calculated using the Iyer-Wald formula \cite{Wald:1993nt}
\begin{equation} \label{Waldentr}
	S=-2\pi \int_{\mathcal{H}} d^2 x \sqrt{\gamma} \mathcal{P}^{\mu \nu \alpha \beta} \epsilon_{\mu \nu} \epsilon_{\alpha \beta}\,,
\end{equation}
with $\mathcal{H}$ being the black hole horizon, $\gamma$ the induced metric determinant at the horizon, $\epsilon_{\mu \nu}$ the anti-symmetric binormal on the horizon surface with normalization $\epsilon_{\mu \nu} \epsilon^{\mu \nu} = -2$, and  $\mathcal{P}^{\mu \nu \rho \sigma}$ is the derivative of the Lagrangian with respect to the Riemann tensor (including the $(16\pi G_{N})^{-1}$ factor), 
\begin{equation}
\mathcal{P}^{\mu \nu \rho \sigma}=\frac{1}{16\pi G_N}\left[g^{\mu[\rho}g^{\sigma]\nu}+P^{\mu \nu \rho \sigma}\right]\, ,
\end{equation}
where $P^{\mu \nu \rho \sigma}$ is the higher-derivative part in \eqref{defofP1}.  For our higher-derivative solution we have 
\begin{align} \label{Waldentrcomp1}
	d^2x\sqrt{\gamma}&=d\theta \sin \theta d\phi \frac{r_+^2}{ K \omega^2} \left( 1 + \frac{1}{2} \lambda_{\rm ev}(f_3(r,\theta)+f_4(r,\theta)) \right)\, ,\\
	\epsilon_{tr} &= \frac{1}{\omega^2} \left( 1 + \frac{1}{2} \lambda_{\rm ev}(f_1(r,\theta)+f_2(r,\theta)) \right)\,,
\end{align}
and the only relevant component of $P^{\mu \nu \rho \sigma}$ is
\begin{align} \label{Waldentrcomp2}
	\mathcal{P}^{trtr} = -\frac{1}{32\pi G_{N}} \omega^4 \left[1 - \lambda_{\rm ev} \left(f_1(r,\theta)+f_2(r,\theta)-\frac{48 m^2 \omega^6}{r^6}+\frac{48 m \omega^3}{r^3L^2}-\frac{12}{L^4}\right) \right].
\end{align}
Thus, integrating \eqref{Waldentr} over $\theta$ and $\phi$ at $r=r_{+}$, we get
\begin{align}
	\frac{S}{S_{\mathrm{GR}}} = 1 + \lambda_{\mathrm{ev}}A^4 \left[\frac{2+10 A^2 r_+^2}{A^4 r_+^4}-\frac{46}{3}-\frac{6 \left(3-7 A^2 r_+^2+A^4 r_+^4\right)}{A^4 L^4 \left(1-A^2 r_+^2\right){}^2}-\frac{72-204 A^2 r_+^2+88 A^4r_+^4}{3 A^4 L^2 r_+^2\left(1- A^2 r_+^2\right)}\right]\, ,
\end{align}
where $S_{\mathrm{GR}}$ is given in \eqref{SGR}.

\subsection*{The missing pieces: mass, string lengths and the first law}
So far, we have determined the temperature, entropy, and string tensions. The last key ingredient to obtain a complete thermodynamic description is the mass $M$ of the black hole. Once the mass is known, we can write down the first law of thermodynamics
\begin{align} \label{eq:firstlaw}
	\delta M = T \delta S + \lambda_{+} \delta \mu_{+} + \lambda_{-} \delta \mu_{-} \,, 
\end{align}
and from it the string lengths $\lambda_{\pm}$ --- conjugate variables to the string tensions --- can be determined.

The computation of the mass is however challenging as the spacetime is not globally asymptotically AdS. A usual approach to obtain the mass of AdS spacetimes is the conformal method of \cite{Ashtekar:1984zz,Ashtekar:2000,Das:2000cu}, known as the AMD method, applied to the C-metric in \cite{Anabalon:2018ydc,Anabalon:2018qfv}. 
However, one cannot apply the AMD formula directly to higher-derivative theories, and obtaining a generalization is not straightforward, especially in the case in which the spacetime is not asymptotically globally AdS. 
The mass can alternatively be derived from a direct computation of the free energy from the Euclidean partition function or from the holographic stress-energy tensor \cite{Anabalon:2018ydc,Anabalon:2018qfv}. However, both of these methods require supplementing the gravitational action with a generalized York-Gibbons-Hawking boundary term \cite{York:1972sj,Gibbons:1976ue} in order to make the variational problem well-posed as well as counterterms in order to remove the divergences. Although effective boundary terms for arbitrary higher-derivative theories are known when the spacetime is asymptotically globally AdS \cite{Bueno:2018xqc}, the fact that the boundary is warped and not globally AdS means that we need fully general boundary terms (valid for arbitrary boundaries and spacetimes). These are known only for some theories like Lovelock gravity \cite{Teitelboim:1987zz,Myers:1987yn} and $f(R)$ gravity \cite{Madsen:1989rz}. To the best of our knowledge such boundary terms have not been explicitly established for the theory \eqref{EFT}\footnote{Nevertheless, the method of auxiliary fields presented in \cite{Deruelle:2009zk} might be useful to this end.}, hence preventing us from computing the mass in this way. 

As a last resource, one can try to derive the value of the mass, along with the string lengths, by assuming the first law. This strategy is a priori valid as independent variations with respect to $A$, $m$ and $K$ yield three equations for three unknowns ($M$, $\lambda_{+}$ and $\lambda_{-}$). However, there is an important nuance: the parameter $\alpha$, which appears in various thermodynamic quantities. 
Although we left it unspecified, this parameter is not free, but should be fixed in terms of the physical parameters of the solution. For solutions that are asymptotically globally AdS, there is a canonical choice for the normalization of the time coordinate by demanding that the speed of light is unity at the boundary. 
However, as the boundary geometry in the case of the accelerating black holes is not maximally symmetric, how to choose the proper normalization is much less clear. The proposal of \cite{Anabalon:2018ydc,Anabalon:2018qfv} is to fix the value of $\alpha$ when $m=0$. In that case, the C-metric in GR becomes AdS in Rindler coordinates, and, as we saw before, performing a change of variables to global AdS leads to the identification \eqref{alphavalue}. The problem is that this identification only works when $m=0$, while $m\neq 0$ deforms the boundary metric and this choice of normalization becomes somewhat unjustified. 

An alternative approach to fix $\alpha$ entails examining the variational problem. As first noted by \cite{Anabalon:2018ydc} and further analyzed by \cite{Kim:2023ncn}, the action does not remain stationary under general variations of the parameters of the solution ($A$, $m$, $K$) unless  $\alpha$ is also varied in a specific way. In particular, it was found by \cite{Kim:2023ncn} that the allowed variations must satisfy 
\begin{align} \label{eq:masterformula}
	0 = \alpha \left(1-2A^2 L^2\right)\delta K-2 K \left(1-A^2 L^2\right)\delta \alpha-2 \alpha  A L^2 K \delta A \,.
\end{align}
This relation fixes $\delta\alpha$ in terms of the other variations. Crucially this relation is not integrable so one cannot obtain a general expression of the form $\alpha(A, m, K)$. It can only be integrated if one restricts the space of variations; for instance \eqref{eq:masterformula} is consistent with \eqref{alphavalue} when $\delta K=0$. However, since in general both expressions are not equivalent when $K$ is varied, these two approaches lead to different values of the string lengths $\lambda_{\pm}$ derived from the first law \cite{Kim:2023ncn}. 
The generalization of \eqref{eq:masterformula} to the case of our higher-derivative solutions requires once again that we supplement the gravitational action with appropriate boundary terms that make the variational problem well-posed. 

To sum up, obtaining the generalized York-Gibbons-Hawking term for \eqref{EFT} is crucial in order to provide a rigorous derivation of the mass, the first law and the string lengths for accelerating black holes in higher-derivative gravity. This problem is beyond the scope of the present paper and we expect to study it elsewhere.

\section{Conclusions}\label{sec:conclusions}
We have found closed-form accelerating black hole solutions in higher-derivative gravity as a perturbative correction to the C-metric. The analysis of these solutions entails many subtleties due to the strange properties of the C-metric. 
First of all, the fact that the C-metric has an exotic asymptotic structure (for instance, in AdS its boundary metric depends on the parameters of the solution) means that we do not have a natural choice for the boundary conditions. As a consequence, there are multiple ways (perhaps, infinitely many) in which the metric can be deformed.
Thus, the solutions that we found contain the higher-derivative corrections (which never vanish) plus extra deformation parameters that one could turn on even at the level of GR. 
A possibility to fix these ambiguities would be to force the corrected geometry to have the same asymptotic metric as the original C-metric. However, a solution satisfying this boundary condition probably does not exist in analytic form, and furthermore, since the C-metric already has an exotic asymptotic structure, there is a priori no reason to exclude additional deformations of the boundary geometry. 
We see no fundamental reason to discard these extra parameters: all of the corresponding solutions we found represent accelerating black holes with the same qualitative features as the C-metric, but different quantitative asymptotic structure.

Our analysis has described both even-parity and odd-parity corrections to GR, and we have observed that they lead to very different behavior. While the even-parity corrections can be captured by an ansatz that preserves the diagonal form of the C-metric \eqref{evenansatz}, odd-parity corrections necessarily introduce an off-diagonal component $dtd\phi$ \eqref{eq:oddansatz}, analogous to the effect of rotation. Even if one sets the angular velocity of the horizon and of the asymptotic frame to zero, the non-vanishing $dtd\phi$ component implies that the spacetime necessarily features rotational frame-dragging at intermediate locations. This is a fascinating example in which linear acceleration plus parity violation lead to rotation. 

We also initiated the analysis of the thermodynamic properties of these black holes in the slowly-accelerating case in AdS ($AL<1$), where the solution represents a single black hole accelerating towards (or from) the AdS boundary. We have obtained the temperature, entropy and string tensions of the solutions, showing that these satisfy some necessary conditions in order to have consistent black hole thermodynamics. We found that the temperature is constant on the horizon, so the zeroth law of black hole mechanics is satisfied, and similarly the string tensions are constant along the axis of acceleration.  However, obtaining the rest of the thermodynamic quantities --- the mass and the string lengths --- is much more subtle. A rigorous derivation of these quantities and of the first law requires a well-posed formulation of the gravitational action, including a generalized York-Gibbons-Hawking term and holographic counterterms. The fact that the boundary is not globally asymptotically AdS makes this problem harder than for black hole solutions with standard asymptotics \cite{Bueno:2018xqc}. 

Besides the completion of the thermodynamic properties and analyzing their holographic interpretation, the solutions presented here have applications in several directions. For instance, in the context of braneworld holography, the C-metric is the cornerstone in the construction of the quantum BTZ black hole. Top-down holographic theories are known to contain an infinite tower of higher-derivative corrections, so our solutions could allow us to determine how these affect the quantum black hole geometries. Other directions entail the analysis of black hole pair-production beyond GR or the construction of more general accelerating black hole solutions in effective field theory, \textit{e.g.}, including electric charge and/or rotation.

\section*{Acknowledgements}
We would like to thank Stéphane Detournay, Tomáš Hale, Robie Hennigar, David Kubiznak, Juan Pedraza, Aaron Poole and Andy Svesko for useful discussions. The work
of P.A.C. is supported by a Ram\'on y Cajal fellowship (RYC2023-044375-I) and by a Proyecto
de generaci\'on de conocimiento (PID2024-155685NB-C22) from Spain’s Ministry of Science,
Innovation and Universities. MD is funded by the Postdoctoral Fellows of the Research Foundation - Flanders grant (1235324N) 
and by the European Union’s Horizon Europe research and innovation programme under the Marie Sk\l{}odowska-Curie grant agreement No. 101206221. This research was carried out in part during a scientific visit funded by the COST action CA22113 "Fundamental Challenges in Theoretical Physics."

\bibliographystyle{JHEP}
\bibliography{Gravities.bib}

\providecommand{\href}[2]{#2}\begingroup\raggedright\begin{thebibliography}{10}

\bibitem{levi-civita:1918}
T.~Levi-Civita, \emph{Ds2 einsteiniani in campi newtoniani. 7. Il sottocaso B2:
  soluzioni oblique : nota del socio T. Levi-Civita}.

\bibitem{Weyl:1917gp}
H.~Weyl, \emph{{The theory of gravitation}},
  \href{http://dx.doi.org/10.1007/s10714-011-1310-7}{\emph{Annalen Phys.}
  {\bfseries 54} (1917) 117--145}.

\bibitem{Ehlers:1962zz}
J.~Ehlers and W.~Kundt, \emph{{Exact solutions of the gravitational field
  equations}}, .

\bibitem{Kinnersley:1970zw}
W.~Kinnersley and M.~Walker, \emph{{Uniformly accelerating charged mass in
  general relativity}},
  \href{http://dx.doi.org/10.1103/PhysRevD.2.1359}{\emph{Phys. Rev. D}
  {\bfseries 2} (1970) 1359--1370}.

\bibitem{Ashtekar:1981ar}
A.~Ashtekar and T.~Dray, \emph{{On the Existence of Solutions to Einstein's
  Equation With Nonzero Bondi News}},
  \href{http://dx.doi.org/10.1007/BF01209313}{\emph{Commun. Math. Phys.}
  {\bfseries 79} (1981) 581--589}.

\bibitem{Plebanski:1976gy}
J.~F. Plebanski and M.~Demianski, \emph{{Rotating, charged, and uniformly
  accelerating mass in general relativity}},
  \href{http://dx.doi.org/10.1016/0003-4916(76)90240-2}{\emph{Annals Phys.}
  {\bfseries 98} (1976) 98--127}.

\bibitem{Hale:2023dpf}
T.~Hale, D.~Kubiz{\v{n}}{\'a}k, O.~Sv{\'\i}tek and T.~Tahamtan,
  \emph{{Solutions and basic properties of regularized Maxwell theory}},
  \href{http://dx.doi.org/10.1103/PhysRevD.107.124031}{\emph{Phys. Rev. D}
  {\bfseries 107} (2023) 124031},
  [\href{https://arxiv.org/abs/2303.16928}{{\ttfamily 2303.16928}}].

\bibitem{Hale:2025veb}
T.~Hale, D.~Kubiz{\v{n}}{\'a}k, J.~Men{\v{s}}{\'\i}kov{\'a}, R.~B. Mann and
  J.~Yang, \emph{{Thermodynamics of charged and accelerating black holes}},
  \href{http://dx.doi.org/10.1103/PhysRevD.111.104004}{\emph{Phys. Rev. D}
  {\bfseries 111} (2025) 104004},
  [\href{https://arxiv.org/abs/2501.13679}{{\ttfamily 2501.13679}}].

\bibitem{EslamPanah:2026iux}
B.~Eslam~Panah, N.~Heidari and {\'A}.~Rinc{\'o}n, \emph{{Accelerating
  electrically charged ModMax black hole solutions in F(R) gravity}},
  \href{http://dx.doi.org/10.1140/epjc/s10052-026-15978-5}{\emph{Eur. Phys. J.
  C} {\bfseries 86} (2026) 705},
  [\href{https://arxiv.org/abs/2607.15914}{{\ttfamily 2607.15914}}].

\bibitem{Anber:2008zz}
M.~M. Anber, \emph{{AdS(4) / CFT(3) + Gravity for Accelerating Conical
  Singularities}},
  \href{http://dx.doi.org/10.1088/1126-6708/2008/11/026}{\emph{JHEP} {\bfseries
  11} (2008) 026}, [\href{https://arxiv.org/abs/0809.2789}{{\ttfamily
  0809.2789}}].

\bibitem{Xu:2011vp}
W.~Xu, K.~Meng and L.~Zhao, \emph{{Accelerating BTZ spacetime}},
  \href{http://dx.doi.org/10.1088/0264-9381/29/15/155005}{\emph{Class. Quant.
  Grav.} {\bfseries 29} (2012) 155005},
  [\href{https://arxiv.org/abs/1111.0730}{{\ttfamily 1111.0730}}].

\bibitem{Astorino:2011mw}
M.~Astorino, \emph{{Accelerating black hole in 2+1 dimensions and 3+1 black
  (st)ring}}, \href{http://dx.doi.org/10.1007/JHEP01(2011)114}{\emph{JHEP}
  {\bfseries 01} (2011) 114},
  [\href{https://arxiv.org/abs/1101.2616}{{\ttfamily 1101.2616}}].

\bibitem{Arenas-Henriquez:2022www}
G.~Arenas-Henriquez, R.~Gregory and A.~Scoins, \emph{{On acceleration in three
  dimensions}}, \href{http://dx.doi.org/10.1007/JHEP05(2022)063}{\emph{JHEP}
  {\bfseries 05} (2022) 063},
  [\href{https://arxiv.org/abs/2202.08823}{{\ttfamily 2202.08823}}].

\bibitem{Cisterna:2023qhh}
A.~Cisterna, F.~Diaz, R.~B. Mann and J.~Oliva, \emph{{Exploring accelerating
  hairy black holes in 2+1 dimensions: the asymptotically locally anti-de
  Sitter class and its holography}},
  \href{http://dx.doi.org/10.1007/JHEP11(2023)073}{\emph{JHEP} {\bfseries 11}
  (2023) 073}, [\href{https://arxiv.org/abs/2309.05559}{{\ttfamily
  2309.05559}}].

\bibitem{Emparan:1995je}
R.~Emparan, \emph{{Pair creation of black holes joined by cosmic strings}},
  \href{http://dx.doi.org/10.1103/PhysRevLett.75.3386}{\emph{Phys. Rev. Lett.}
  {\bfseries 75} (1995) 3386--3389},
  [\href{https://arxiv.org/abs/gr-qc/9506025}{{\ttfamily gr-qc/9506025}}].

\bibitem{Emparan:1995eb}
R.~Emparan, \emph{{Correlations between black holes formed in cosmic string
  breaking}}, \href{http://dx.doi.org/10.1103/PhysRevD.52.6976}{\emph{Phys.
  Rev. D} {\bfseries 52} (1995) 6976--6981},
  [\href{https://arxiv.org/abs/gr-qc/9507002}{{\ttfamily gr-qc/9507002}}].

\bibitem{Hawking:1995zn}
S.~W. Hawking and S.~F. Ross, \emph{{Pair production of black holes on cosmic
  strings}}, \href{http://dx.doi.org/10.1103/PhysRevLett.75.3382}{\emph{Phys.
  Rev. Lett.} {\bfseries 75} (1995) 3382--3385},
  [\href{https://arxiv.org/abs/gr-qc/9506020}{{\ttfamily gr-qc/9506020}}].

\bibitem{Mann:1996gj}
R.~B. Mann, \emph{{Pair production of topological anti-de Sitter black holes}},
  \href{http://dx.doi.org/10.1088/0264-9381/14/5/007}{\emph{Class. Quant.
  Grav.} {\bfseries 14} (1997) L109--L114},
  [\href{https://arxiv.org/abs/gr-qc/9607071}{{\ttfamily gr-qc/9607071}}].

\bibitem{Booth:1998gf}
I.~S. Booth and R.~B. Mann, \emph{{Cosmological pair production of charged and
  rotating black holes}},
  \href{http://dx.doi.org/10.1016/S0550-3213(98)00756-1}{\emph{Nucl. Phys. B}
  {\bfseries 539} (1999) 267--306},
  [\href{https://arxiv.org/abs/gr-qc/9806056}{{\ttfamily gr-qc/9806056}}].

\bibitem{Dias:2003st}
O.~J.~C. Dias and J.~P.~S. Lemos, \emph{{Pair creation of de Sitter black holes
  on a cosmic string background}},
  \href{http://dx.doi.org/10.1103/PhysRevD.69.084006}{\emph{Phys. Rev. D}
  {\bfseries 69} (2004) 084006},
  [\href{https://arxiv.org/abs/hep-th/0310068}{{\ttfamily hep-th/0310068}}].

\bibitem{Dias:2004rz}
O.~J.~C. Dias, \emph{{Pair creation of anti-de Sitter black holes on a cosmic
  string background}},
  \href{http://dx.doi.org/10.1103/PhysRevD.70.024007}{\emph{Phys. Rev. D}
  {\bfseries 70} (2004) 024007},
  [\href{https://arxiv.org/abs/hep-th/0401069}{{\ttfamily hep-th/0401069}}].

\bibitem{Emparan:1999fd}
R.~Emparan, G.~T. Horowitz and R.~C. Myers, \emph{{Exact description of black
  holes on branes. 2. Comparison with BTZ black holes and black strings}},
  \href{http://dx.doi.org/10.1088/1126-6708/2000/01/021}{\emph{JHEP} {\bfseries
  01} (2000) 021}, [\href{https://arxiv.org/abs/hep-th/9912135}{{\ttfamily
  hep-th/9912135}}].

\bibitem{Emparan:2000fn}
R.~Emparan, R.~Gregory and C.~Santos, \emph{{Black holes on thick branes}},
  \href{http://dx.doi.org/10.1103/PhysRevD.63.104022}{\emph{Phys. Rev. D}
  {\bfseries 63} (2001) 104022},
  [\href{https://arxiv.org/abs/hep-th/0012100}{{\ttfamily hep-th/0012100}}].

\bibitem{Gregory:2008br}
R.~Gregory, S.~F. Ross and R.~Zegers, \emph{{Classical and quantum gravity of
  brane black holes}},
  \href{http://dx.doi.org/10.1088/1126-6708/2008/09/029}{\emph{JHEP} {\bfseries
  09} (2008) 029}, [\href{https://arxiv.org/abs/0802.2037}{{\ttfamily
  0802.2037}}].

\bibitem{Emparan:2020znc}
R.~Emparan, A.~M. Frassino and B.~Way, \emph{{Quantum BTZ black hole}},
  \href{http://dx.doi.org/10.1007/JHEP11(2020)137}{\emph{JHEP} {\bfseries 11}
  (2020) 137}, [\href{https://arxiv.org/abs/2007.15999}{{\ttfamily
  2007.15999}}].

\bibitem{Climent:2024nuj}
A.~Climent, R.~Emparan and R.~A. Hennigar, \emph{{Chemical potential and charge
  in quantum black holes}},
  \href{http://dx.doi.org/10.1007/JHEP08(2024)150}{\emph{JHEP} {\bfseries 08}
  (2024) 150}, [\href{https://arxiv.org/abs/2404.15148}{{\ttfamily
  2404.15148}}].

\bibitem{Bhattacharya:2025tdn}
D.~Bhattacharya, R.~A. Hennigar, R.~B. Mann and M.~Zhang, \emph{{Charged
  rotating quantum black holes}},
  \href{http://dx.doi.org/10.1007/JHEP11(2025)165}{\emph{JHEP} {\bfseries 11}
  (2025) 165}, [\href{https://arxiv.org/abs/2506.19941}{{\ttfamily
  2506.19941}}].

\bibitem{Astorino:2016ybm}
M.~Astorino, \emph{{Thermodynamics of Regular Accelerating Black Holes}},
  \href{http://dx.doi.org/10.1103/PhysRevD.95.064007}{\emph{Phys. Rev. D}
  {\bfseries 95} (2017) 064007},
  [\href{https://arxiv.org/abs/1612.04387}{{\ttfamily 1612.04387}}].

\bibitem{Appels:2016uha}
M.~Appels, R.~Gregory and D.~Kubiznak, \emph{{Thermodynamics of Accelerating
  Black Holes}},
  \href{http://dx.doi.org/10.1103/PhysRevLett.117.131303}{\emph{Phys. Rev.
  Lett.} {\bfseries 117} (2016) 131303},
  [\href{https://arxiv.org/abs/1604.08812}{{\ttfamily 1604.08812}}].

\bibitem{Appels:2017xoe}
M.~Appels, R.~Gregory and D.~Kubiznak, \emph{{Black Hole Thermodynamics with
  Conical Defects}},
  \href{http://dx.doi.org/10.1007/JHEP05(2017)116}{\emph{JHEP} {\bfseries 05}
  (2017) 116}, [\href{https://arxiv.org/abs/1702.00490}{{\ttfamily
  1702.00490}}].

\bibitem{Gregory:2017ogk}
R.~Gregory, \emph{{Accelerating Black Holes}},
  \href{http://dx.doi.org/10.1088/1742-6596/942/1/012002}{\emph{J. Phys. Conf.
  Ser.} {\bfseries 942} (2017) 012002},
  [\href{https://arxiv.org/abs/1712.04992}{{\ttfamily 1712.04992}}].

\bibitem{Anabalon:2018qfv}
A.~Anabal{\'o}n, F.~Gray, R.~Gregory, D.~Kubiz{\v{n}}{\'a}k and R.~B. Mann,
  \emph{{Thermodynamics of Charged, Rotating, and Accelerating Black Holes}},
  \href{http://dx.doi.org/10.1007/JHEP04(2019)096}{\emph{JHEP} {\bfseries 04}
  (2019) 096}, [\href{https://arxiv.org/abs/1811.04936}{{\ttfamily
  1811.04936}}].

\bibitem{Anabalon:2018ydc}
A.~Anabal{\'o}n, M.~Appels, R.~Gregory, D.~Kubiz{\v{n}}{\'a}k, R.~B. Mann and
  A.~Ovg{\"u}n, \emph{{Holographic Thermodynamics of Accelerating Black
  Holes}}, \href{http://dx.doi.org/10.1103/PhysRevD.98.104038}{\emph{Phys. Rev.
  D} {\bfseries 98} (2018) 104038},
  [\href{https://arxiv.org/abs/1805.02687}{{\ttfamily 1805.02687}}].

\bibitem{Kim:2023ncn}
H.~Kim, N.~Kim, Y.~Lee and A.~Poole, \emph{{Thermodynamics of accelerating
  AdS$_4$ black holes from the covariant phase space}},
  \href{http://dx.doi.org/10.1140/epjc/s10052-023-12266-4}{\emph{Eur. Phys. J.
  C} {\bfseries 83} (2023) 1095},
  [\href{https://arxiv.org/abs/2306.16187}{{\ttfamily 2306.16187}}].

\bibitem{Hubeny:2009kz}
V.~E. Hubeny, D.~Marolf and M.~Rangamani, \emph{{Black funnels and droplets
  from the AdS C-metrics}},
  \href{http://dx.doi.org/10.1088/0264-9381/27/2/025001}{\emph{Class. Quant.
  Grav.} {\bfseries 27} (2010) 025001},
  [\href{https://arxiv.org/abs/0909.0005}{{\ttfamily 0909.0005}}].

\bibitem{Astorino:2016xiy}
M.~Astorino, \emph{{CFT Duals for Accelerating Black Holes}},
  \href{http://dx.doi.org/10.1016/j.physletb.2016.07.019}{\emph{Phys. Lett. B}
  {\bfseries 760} (2016) 393--405},
  [\href{https://arxiv.org/abs/1605.06131}{{\ttfamily 1605.06131}}].

\bibitem{Arenas-Henriquez:2023hur}
G.~Arenas-Henriquez, A.~Cisterna, F.~Diaz and R.~Gregory, \emph{{Accelerating
  Black Holes in $2+1$ dimensions: Holography revisited}},
  \href{http://dx.doi.org/10.1007/JHEP09(2023)122}{\emph{JHEP} {\bfseries 09}
  (2023) 122}, [\href{https://arxiv.org/abs/2308.00613}{{\ttfamily
  2308.00613}}].

\bibitem{Arenas-Henriquez:2025rpt}
G.~Arenas-Henriquez, L.~Ciambelli, F.~Diaz, W.~Jia and D.~Rivera-Betancour,
  \emph{{Radiation in fluid/gravity and the flat limit}},
  \href{http://dx.doi.org/10.1007/JHEP01(2026)086}{\emph{JHEP} {\bfseries 01}
  (2026) 086}, [\href{https://arxiv.org/abs/2508.01446}{{\ttfamily
  2508.01446}}].

\bibitem{Lu:2014sza}
H.~L{\"u} and J.~F. V{\'a}zquez-Poritz, \emph{{C-metrics in Gauged STU
  Supergravity and Beyond}},
  \href{http://dx.doi.org/10.1007/JHEP12(2014)057}{\emph{JHEP} {\bfseries 12}
  (2014) 057}, [\href{https://arxiv.org/abs/1408.6531}{{\ttfamily 1408.6531}}].

\bibitem{Ferrero:2020twa}
P.~Ferrero, J.~P. Gauntlett, J.~M.~P. Ipi{\~n}a, D.~Martelli and J.~Sparks,
  \emph{{Accelerating black holes and spinning spindles}},
  \href{http://dx.doi.org/10.1103/PhysRevD.104.046007}{\emph{Phys. Rev. D}
  {\bfseries 104} (2021) 046007},
  [\href{https://arxiv.org/abs/2012.08530}{{\ttfamily 2012.08530}}].

\bibitem{Cassani:2021dwa}
D.~Cassani, J.~P. Gauntlett, D.~Martelli and J.~Sparks, \emph{{Thermodynamics
  of accelerating and supersymmetric AdS4 black holes}},
  \href{http://dx.doi.org/10.1103/PhysRevD.104.086005}{\emph{Phys. Rev. D}
  {\bfseries 104} (2021) 086005},
  [\href{https://arxiv.org/abs/2106.05571}{{\ttfamily 2106.05571}}].

\bibitem{Ferrero:2021ovq}
P.~Ferrero, M.~Inglese, D.~Martelli and J.~Sparks, \emph{{Multicharge
  accelerating black holes and spinning spindles}},
  \href{http://dx.doi.org/10.1103/PhysRevD.105.126001}{\emph{Phys. Rev. D}
  {\bfseries 105} (2022) 126001},
  [\href{https://arxiv.org/abs/2109.14625}{{\ttfamily 2109.14625}}].

\bibitem{Ball:2020vzo}
A.~Ball and N.~Miller, \emph{{Accelerating black hole thermodynamics with boost
  time}}, \href{http://dx.doi.org/10.1088/1361-6382/ac0766}{\emph{Class. Quant.
  Grav.} {\bfseries 38} (2021) 145031},
  [\href{https://arxiv.org/abs/2008.03682}{{\ttfamily 2008.03682}}].

\bibitem{Ball:2021xwt}
A.~Ball, \emph{{Global first laws of accelerating black holes}},
  \href{http://dx.doi.org/10.1088/1361-6382/ac2139}{\emph{Class. Quant. Grav.}
  {\bfseries 38} (2021) 195024},
  [\href{https://arxiv.org/abs/2103.07521}{{\ttfamily 2103.07521}}].

\bibitem{Podolsky:2002nk}
J.~Podolsky, \emph{{Accelerating black holes in anti-de Sitter universe}},
  \href{http://dx.doi.org/10.1023/A:1013961411430}{\emph{Czech. J. Phys.}
  {\bfseries 52} (2002) 1--10},
  [\href{https://arxiv.org/abs/gr-qc/0202033}{{\ttfamily gr-qc/0202033}}].

\bibitem{Dias:2002mi}
O.~J.~C. Dias and J.~P.~S. Lemos, \emph{{Pair of accelerated black holes in
  anti-de Sitter background: AdS C metric}},
  \href{http://dx.doi.org/10.1103/PhysRevD.67.064001}{\emph{Phys. Rev. D}
  {\bfseries 67} (2003) 064001},
  [\href{https://arxiv.org/abs/hep-th/0210065}{{\ttfamily hep-th/0210065}}].

\bibitem{Krtous:2005ej}
P.~Krtous, \emph{{Accelerated black holes in an anti-de Sitter universe}},
  \href{http://dx.doi.org/10.1103/PhysRevD.72.124019}{\emph{Phys. Rev. D}
  {\bfseries 72} (2005) 124019},
  [\href{https://arxiv.org/abs/gr-qc/0510101}{{\ttfamily gr-qc/0510101}}].

\bibitem{Meng:2016gyt}
K.~Meng and L.~Zhao, \emph{{C-metric solution for conformal gravity with a
  conformally coupled scalar field}},
  \href{http://dx.doi.org/10.1016/j.aop.2017.01.001}{\emph{Annals Phys.}
  {\bfseries 377} (2017) 466--483},
  [\href{https://arxiv.org/abs/1601.07634}{{\ttfamily 1601.07634}}].

\bibitem{Lim:2016lxk}
Y.-K. Lim, \emph{{Charged C-metric in conformal gravity}},
  \href{http://dx.doi.org/10.1103/PhysRevD.93.084045}{\emph{Phys. Rev. D}
  {\bfseries 93} (2016) 084045},
  [\href{https://arxiv.org/abs/1604.07485}{{\ttfamily 1604.07485}}].

\bibitem{Zhang:2019vpf}
M.~Zhang and R.~B. Mann, \emph{{Charged accelerating black hole in $f(R)$
  gravity}}, \href{http://dx.doi.org/10.1103/PhysRevD.100.084061}{\emph{Phys.
  Rev. D} {\bfseries 100} (2019) 084061},
  [\href{https://arxiv.org/abs/1908.05118}{{\ttfamily 1908.05118}}].

\bibitem{Cisterna:2021xxq}
A.~Cisterna, A.~Neira-Gallegos, J.~Oliva and S.~C. Rebolledo-Caceres,
  \emph{{Pleba{\'n}ski-Demia{\'n}ski solutions in quadratic gravity with
  conformally coupled scalar fields}},
  \href{http://dx.doi.org/10.1103/PhysRevD.103.064050}{\emph{Phys. Rev. D}
  {\bfseries 103} (2021) 064050},
  [\href{https://arxiv.org/abs/2101.03628}{{\ttfamily 2101.03628}}].

\bibitem{Suryaatmadja:2026ais}
C.~Suryaatmadja, D.-h. Yeom and R.~Mann, \emph{{Acceleration in 3D
  Einstein-Gauss-Bonnet Gravity}},
  \href{https://arxiv.org/abs/2607.08030}{{\ttfamily 2607.08030}}.

\bibitem{Hong:2003gx}
K.~Hong and E.~Teo, \emph{{A New form of the C metric}},
  \href{http://dx.doi.org/10.1088/0264-9381/20/14/321}{\emph{Class. Quant.
  Grav.} {\bfseries 20} (2003) 3269--3277},
  [\href{https://arxiv.org/abs/gr-qc/0305089}{{\ttfamily gr-qc/0305089}}].

\bibitem{Griffiths:2006tk}
J.~B. Griffiths, P.~Krtous and J.~Podolsky, \emph{{Interpreting the C-metric}},
  \href{http://dx.doi.org/10.1088/0264-9381/23/23/008}{\emph{Class. Quant.
  Grav.} {\bfseries 23} (2006) 6745--6766},
  [\href{https://arxiv.org/abs/gr-qc/0609056}{{\ttfamily gr-qc/0609056}}].

\bibitem{Smarr:1973zz}
L.~Smarr, \emph{{Surface Geometry of Charged Rotating Black Holes}},
  \href{http://dx.doi.org/10.1103/PhysRevD.7.289}{\emph{Phys. Rev. D}
  {\bfseries 7} (1973) 289--295}.

\bibitem{Cano:2019ore}
P.~A. Cano and A.~Ruip\'erez, \emph{{Leading higher-derivative corrections to
  Kerr geometry}}, \href{http://dx.doi.org/10.1007/JHEP05(2019)189}{\emph{JHEP}
  {\bfseries 05} (2019) 189},
  [\href{https://arxiv.org/abs/1901.01315}{{\ttfamily 1901.01315}}].

\bibitem{Wald:1993nt}
R.~M. Wald, \emph{{Black hole entropy is the Noether charge}},
  \href{http://dx.doi.org/10.1103/PhysRevD.48.R3427}{\emph{Phys. Rev. D}
  {\bfseries 48} (1993) R3427--R3431},
  [\href{https://arxiv.org/abs/gr-qc/9307038}{{\ttfamily gr-qc/9307038}}].

\bibitem{Ashtekar:1984zz}
A.~Ashtekar and A.~Magnon, \emph{{Asymptotically anti-de Sitter space-times}},
  \href{http://dx.doi.org/10.1088/0264-9381/1/4/002}{\emph{Class. Quant. Grav.}
  {\bfseries 1} (1984) L39--L44}.

\bibitem{Ashtekar:2000}
A.~Ashtekar and S.~Das, \emph{Asymptotically anti-de sitter space-times:
  Conserved quantities},
  \href{https://arxiv.org/abs/hep-th/9911230}{{\ttfamily hep-th/9911230}}.

\bibitem{Das:2000cu}
S.~Das and R.~B. Mann, \emph{{Conserved quantities in Kerr-anti-de Sitter
  space-times in various dimensions}},
  \href{http://dx.doi.org/10.1088/1126-6708/2000/08/033}{\emph{JHEP} {\bfseries
  08} (2000) 033}, [\href{https://arxiv.org/abs/hep-th/0008028}{{\ttfamily
  hep-th/0008028}}].

\bibitem{York:1972sj}
J.~W. York, Jr., \emph{{Role of conformal three geometry in the dynamics of
  gravitation}},
  \href{http://dx.doi.org/10.1103/PhysRevLett.28.1082}{\emph{Phys. Rev. Lett.}
  {\bfseries 28} (1972) 1082--1085}.

\bibitem{Gibbons:1976ue}
G.~W. Gibbons and S.~W. Hawking, \emph{{Action Integrals and Partition
  Functions in Quantum Gravity}},
  \href{http://dx.doi.org/10.1103/PhysRevD.15.2752}{\emph{Phys. Rev. D}
  {\bfseries 15} (1977) 2752--2756}.

\bibitem{Bueno:2018xqc}
P.~Bueno, P.~A. Cano and A.~Ruip{\'e}rez, \emph{{Holographic studies of
  Einsteinian cubic gravity}},
  \href{http://dx.doi.org/10.1007/JHEP03(2018)150}{\emph{JHEP} {\bfseries 03}
  (2018) 150}, [\href{https://arxiv.org/abs/1802.00018}{{\ttfamily
  1802.00018}}].

\bibitem{Teitelboim:1987zz}
C.~Teitelboim and J.~Zanelli, \emph{{Dimensionally continued topological
  gravitation theory in Hamiltonian form}},
  \href{http://dx.doi.org/10.1088/0264-9381/4/4/010}{\emph{Class. Quant. Grav.}
  {\bfseries 4} (1987) L125}.

\bibitem{Myers:1987yn}
R.~C. Myers, \emph{{Higher Derivative Gravity, Surface Terms and String
  Theory}}, \href{http://dx.doi.org/10.1103/PhysRevD.36.392}{\emph{Phys. Rev.
  D} {\bfseries 36} (1987) 392}.

\bibitem{Madsen:1989rz}
M.~S. Madsen and J.~D. Barrow, \emph{{De Sitter Ground States and Boundary
  Terms in Generalized Gravity}},
  \href{http://dx.doi.org/10.1016/0550-3213(89)90596-8}{\emph{Nucl. Phys. B}
  {\bfseries 323} (1989) 242--252}.

\bibitem{Deruelle:2009zk}
N.~Deruelle, M.~Sasaki, Y.~Sendouda and D.~Yamauchi, \emph{{Hamiltonian
  formulation of f(Riemann) theories of gravity}},
  \href{http://dx.doi.org/10.1143/PTP.123.169}{\emph{Prog. Theor. Phys.}
  {\bfseries 123} (2010) 169--185},
  [\href{https://arxiv.org/abs/0908.0679}{{\ttfamily 0908.0679}}].

\end{thebibliography}\endgroup

\end{document}